\documentclass[twocolumn]{aastex631}
\usepackage{graphicx,lipsum, booktabs, footnote}
\usepackage{wrapfig}
\usepackage{float}
\usepackage{multirow}

\usepackage[autostyle]{csquotes}

\begin{document}

\title{Globular Clusters GMRT Pulsar Search (GCGPS) I: Survey description, discovery and timing of the first pulsar in NGC~6093 (M80)}

\author[0009-0006-7995-5871]{Jyotirmoy Das}
\author[0000-0002-2892-8025]{Jayanta Roy}
\affiliation{National Centre for Radio Astrophysics (NCRA) \\
Pune-411007, Maharashtra, India}

\author[0000-0003-1307-9435]{Paulo C. C. Freire}
\affiliation{Max-Planck-Institut für Radioastronomie (MPIfR) \\
 Auf dem Hügel 69-53121, Bonn, Germany}
 
\author[0000-0001-5799-9714]{Scott M Ransom}
\affiliation{National Radio Astronomy Observatory (NRAO) \\
 Charlottesville, Virginia, United States}

\author[0000-0002-6287-6900]{Bhaswati Bhattacharyya}
\affiliation{National Centre for Radio Astrophysics (NCRA) \\
Pune-411007, Maharashtra, India}

\author[0000-0003-2797-0595]{Karel Adámek}
\author[0000-0003-1756-3064]{Wes Armour}
\affiliation{Oxford e-Research Centre (OeRC)
University of Oxford, Oxford-OX13PJ, United Kingdom}

\author[0000-0002-6631-1077]{Sanjay Kudale}
\affiliation{National Centre for Radio Astrophysics (NCRA) \\
Pune-411007, Maharashtra, India}
\affiliation{Giant Metrewave Radio Telescope (GMRT) \\
Khodad, Pune- 410504, Maharashtra, India}

\author[0009-0008-1233-6915]{Mekhala V. Muley}
\affiliation{Giant Metrewave Radio Telescope (GMRT) \\
Khodad, Pune- 410504, Maharashtra, India}

\begin{abstract}

This paper describes the new Globular Clusters GMRT Pulsar Search (GCGPS) survey. This survey aims to find Millisecond Pulsars (MSPs) in the Globular Clusters (GCs) of the Milky Way using the upgraded Giant Metrewave Radio Telescope (uGMRT). Using uGMRT's Band-4 (550$–$750 MHz) and Band-3 (300$–$500 MHz) receivers, this survey will eventually cover the GCs accessible to the uGMRT sky, avoiding GCs visible to the Five-hundred-meter Aperture Spherical radio Telescope (FAST) (i.e. $-53^\circ\,<\delta\,<\,-\:17^\circ$), and targeting the GCs that have not been targeted with the sensitivity of this survey. In this paper, we present the discovery and follow-up study of the first pulsar from this survey, J1617$-$2258A, a 4.32 ms binary MSP, the first to be discovered in the GC NGC~6093. We localised this MSP with arc-sec precision from imaging and obtained the unique timing solution from more than one year of timing observations with the uGMRT Band-4 (550$-$750 MHz) receivers. This revealed an unusual binary MSP, with a $\sim$ 19-hour, highly eccentric (e $\sim$ 0.54) orbit having a low-mass companion. This orbital eccentricity allowed the measurement of the rate of advance of periastron for this system, which led to the derivation of its total mass, $1.67 \, \pm \, 0.06 \, \rm M_{\odot}$; this together with the system's mass function implies, for the pulsar and the companion, $M_\mathrm{p} < 1.60 \, \rm M_{\odot}$ and $M_\mathrm{c} > 0.072 \, \rm M_{\odot}$. The system is most likely a perturbed MSP - Helium white dwarf system seen at a low orbital inclination.
\end{abstract}


\section{Introduction}
\label{sec:intro}
The close interaction of an old neutron star (NS) with a low-mass star results in the formation of low-mass Xray-binaries (LMXBs). In such systems, the NS accretes matter from its companion star and gradually spins up.  The high stellar density and high stellar interaction rates \citep{Verbunt_Hut_1987, Pooley_2003, Bahramian_2013} found in Globular Clusters (GCs) allow for the dynamical formation of LMXBs, which is not possible in the Galactic disk, and this is the reason why there are $\sim 10^3$ times more LMXBs per unit mass in GCs compared to the Galactic disk \citep{Sarazin_2003}. At the end of the dynamical evolution of these LMXBs, we are left with pulsars with spin periods of a few milliseconds, called recycled or millisecond pulsars \citep[MSP, defined here as having $P < 20 \rm \, ms$ and a low B-field,][]{Becker_1999}, and a low-mass non-degenerate companion or a low-mass white-dwarf \citep[for a review, see][]{Tauris_2023}. The large number of LMXBs results in a large number of MSPs: out of the 729 discovered MSPs, 311 are found in GCs\footnote{Data taken from the ATNF Pulsar catalogue (\url{https://www.atnf.csiro.au/research/pulsar/psrcat/}) and Paulo Freire's GC website (\url{https://www3.mpifr-bonn.mpg.de/staff/pfreire/GCpsr.html}).}.

The local environment of a GC plays an important role, not only in the formation of LMXBs and MSPs but also afterwards. If the stellar interaction rate per binary, $\gamma$ \citep{Verbunt_Freire_2014} is not very high, the evolution from an LMXB to an MSP is likely to be undisturbed and as a result we are likely to get an MSP with a non-degenerate star as a companion interacting with the MSP (i.e spider MSP), or an MSP in a low-eccentricity orbit with a degenerate companion like a white dwarf (WD), a NS-WD system. These nearly circular systems are indistinguishable from the MSPs in the galactic disc, as observed in lower density GCs, like M3 \citep{Li_2024}, M5 \citep{Zhang_2023}, M13 \citep{Wang_2020} and M53 \citep{Lian_2023, Lian_2025}.

If $\gamma$ is very high, then LMXBs might be disrupted to form exotic MSPs, where the end of the accretion results in the formation of mildly recycled, high-B field MSPs \citep{Freire_2011, Johnson_2013}. But even after the accretion ends, interaction with a third stellar object may lead either to the dissociation of binary systems and the isolation of many pulsars, as observed in core-collapsed GCs like Terzan 1 \citep{Singleton_2024}, NGC~6517 \citep{Lynch2011, Yin_2024}, NGC~6522 \citep{Abbate_2023}, NGC~6624 \citep{Lynch2012, Abbate2022}, NGC~6752 \citep{Corongiu_2024} and M15 \citep{Zhou_2024, Wu_2024}. Another event that becomes more likely is the exchange of the low-mass companion of an MSP, generally for a more massive one, in what has been designated as a {\em secondary exchange}. This produces systems like e.g. PSR~J1807$-$2500B in NGC~6544 \citealt{Lynch2012}, PSR~J1823$-$3021G in NGC~6624, \citealt{Ridolfi2021} and PSR~J1835$-$3259A in NGC 6652, \citealt{DeCesar_2015}, which have no counterpart in the Galactic disk.

For GCs with intermediate densities, like NGC~1851 \citep{Freire2004, Ridolfi2022}, Terzan 5 \citep{Ransom2005, Padmanabh_2024} and M28 \citep{Bogdanov_2011, Douglas_2022}, we see a mix of both types of populations.
Interestingly, some reasonably dense GCs like 47~Tuc \citep{Freire_2017} or M62 \citep{Vleeschower_2024} show a \enquote{normal} MSP population, only truncated in the orbital period.

The large eccentricities and companion masses of some of the exotic systems make them potentially useful probes of gravitational effects in strong field regimes: for instance, NGC 1851E might be an MSP-black hole system \citep{Barr_2024}, which would enable tests of new types of gravity theories that until now could not be tested with binary pulsar timing \citep{Freire_Wex_2024}. Such unusual systems provide a strong motivation for finding additional pulsars in GCs.

Because typical GC distances are of the order of tens of kiloparsecs (kpc), most pulsars, which are generally faint radio sources, remain undetected. Indeed, Galactic disk MSPs are typically found within distances of the order of a kpc. This implies two things: 1) the vast majority of GC pulsars remain undetected, meaning many more pulsars in GCs are yet to be discovered; 2) discovering them will require more sensitive surveys.

Most of the successful and most recent surveys have been conducted in the GHz frequency range, either in L-Band, like Parkes and Arecibo GC surveys \citep[carried out around 2000,][]{Camilo2000, Possenti2001, Possenti2005, Amico2003, Hessels_2007},
the ongoing Transients and Pulsars with MeerKAT \citep[TRAPUM][]{Ridolfi2021, Ridolfi2022} and FAST \citep{Pan2021a, Pan2021b} surveys
or in S-Band, like the GBT S-Band GC Pulsar Survey (see \citealt{Ransom2004, Ransom2005, Freire2008, Lynch2011}, see also discussion in Section \ref{sec:motivation}); this means that such surveys are biased towards relatively flat-spectrum MSPs. The advent of the upgraded Giant Metrewave Radio Telescope (uGMRT) and its seamless low-frequency coverage (300$-$850 $\rm MHz$ for uGMRT Band-3 and Band-4) allowed us to use its Y-shaped array and effectively design a pulsar survey to search for GC MSPs in the low-frequency range. Searching GC MSPs with the uGMRT will allow us to discover steep-spectrum GC MSPs that may have been missed earlier due to the bias towards flat-spectrum MSPs in surveys at higher frequencies.

In this paper, we discuss the Globular Clusters GMRT Pulsar Search (GCGPS), from its motivation (Section \ref{sec:motivation}) to its design, sensitivity, and target selection in Section \ref{sec:survey}. In Section \ref{sec:dataanalysis}, a detailed discussion is done on the strategy of data analysis. In Section \ref{subsection:discovery} we report the first GCGPS discovery, a pulsar in the GC NGC 6093 (also known as Messier 80, henceforth M80), followed by its localization in Section \ref{subsec:localization}, profile study in Section \ref{subsec:profile}, timing in Section \ref{subsec:timing} followed by its binary properties in Section \ref{subsec:characteristics} and finally we summarize our work in Section \ref{sec:summary}.

\section{MOTIVATION FOR DESIGNING AN GC MSP SEARCH SURVEY with the uGMRT}
\label{sec:motivation}

The improvement in sensitivity achieved with the uGMRT relative to the legacy GMRT implies the possibility of additional pulsar discoveries. Furthermore, the large number of undiscovered pulsars in GCs and the possibility of finding exotic binaries there that are unlike the Galactic disk pulsars serve as the motivation for this survey.

There are two previously reported GC MSP discoveries with GMRT. A survey of 10 GCs was made with the legacy GMRT, which resulted in the discovery of the first MSP in NGC~1851, PSR~J0514$-$4002A, a steep-spectrum 5-ms pulsar that is a member of one of the first exotic binaries known that was clearly formed in a secondary exchange encounter \citep{Freire2004}. The most recent effort was a shallow survey made by \cite{Gautam2022}, where they discovered the second MSP of NGC~6652, PSR~J1835$-$3259B, an exceptionally young, powerful and radio bright system that has a bright, young WD companion \citep{Chen_2023} and is one of the three GC pulsars detected in $\gamma$-rays \citep{Zhang_2022}. These surveys, but especially the latter, take full advantage of the imaging capabilities of the uGMRT, which have uncovered additional candidates, both in pulsation and imaging data.

One crucial factor about the existing and ongoing searches of pulsars from the GCs is that they mostly focus on the GCs that have discovered pulsars in them, as the known DM reduces the computational requirements. MeerKAT's TRAPUM and the uGMRT survey by \cite{Gautam2022} followed this strategy. Because of this several GCs with no discovered MSPs remained unobserved by sensitive surveys like MeerKAT's TRAPUM or uGMRT; even though GBT or Parkes previously observed some of the GCs, those observations were mostly in GHz frequency \citep{Ransom2004, Ransom2005, Possenti2005} and were significantly less sensitive compared to uGMRT Band-4 (550$-$750 MHz) central square phased array beam sensitivity for the same on-source time (discussed in Section \ref{subsec:target}), for steep spectrum pulsars. uGMRT's distinctive low-frequency (300$-$850 $\rm MHz$ for uGMRT Band-3 and Band-4) coverage, particularly Band-4 (550$-$750 $\rm MHz$) as discussed in Section \ref{subsec:design}, makes the GCs that may have never been searched or missed by earlier high-frequency surveys promising targets to search for new MSPs.

Due to FAST and MeerKAT's superior sensitivity (quantified in Section \ref{subsec:target}), the GCs that remained unobserved by FAST's GC MSP search (L-Band), or MeerKAT's TRAPUM (UHF and L-Band survey), make a good set of GCs to design an effective survey to search for MSPs with uGMRT. After careful consideration, we got a significant number of such observable GCs and the majority of them met our additional selection criteria (discussed in Section \ref{subsec:target}), which motivated us to design the GCGPS\footnote{Survey webpage: \url{http://www.ncra.tifr.res.in/~jroy/GC.html}} survey.

\section{The survey}
\label{sec:survey}

In this section, we will discuss the GCGPS survey details, from strategic design to the theoretical estimation of the survey sensitivity and target selection.

\subsection{Designing the survey}\label{subsec:design}

\begin{figure}
\centering
    \includegraphics[width=\columnwidth]{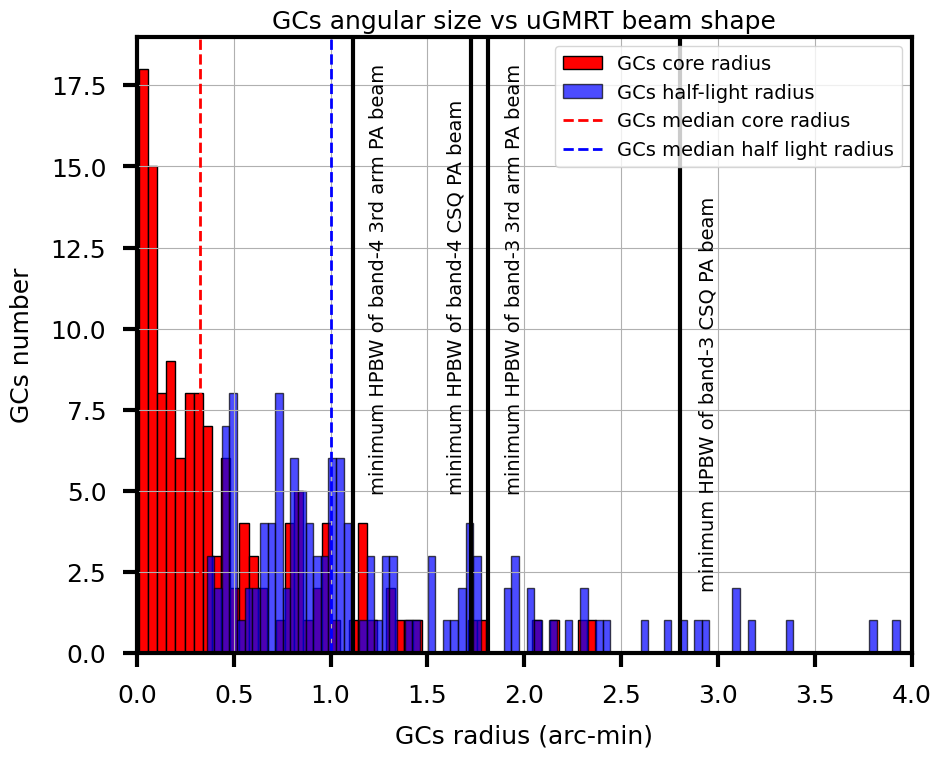}
    \caption{Histogram plot of all the GCs core and half-light radius. The vertically dotted red line represents the median core radius value and the similar blue line represents the median half-light radius of all the GCs. All the GC data was taken from W. E. Harris's GC catalogue\citep[2010 edition]{Harris_1996}. The four vertical solid black lines represent the uGMRT Band-3 and Band-4, CSQ and 3rd arm PA beam's minimum Half Power Bandwidth ($HPBW_{min}$), as marked in the plot.}
    \label{fig:GCradiusvsbeamsize}
\end{figure}

The uGMRT has 30 antennae distributed in a Y-shaped fashion, with a compact array of 14 antennae at the centre within a $1 \, \rm km^2$ area, and the other 16 antennae are distributed in the three arms with a maximum baseline length of $\sim 28 \rm \, km$ to make the array suitable to observe compact objects \citep{Swarup_1991, Gupta_2017}.

To design this GC MSP search survey effectively for uGMRT, we need to have an initial estimation of the typical target size for this survey. The known Milky Way GCs have a median core and half-light radius of 0.325 and 1.0 arcminutes respectively (see Figure \ref{fig:GCradiusvsbeamsize}). With this estimation, to select our optimal uGMRT observing Band and beam type, using a uGMRT beam simulation code {\tt uGMRT\_beam\_fov.py}\footnote{See \url{https://www.ursi.org/proceedings/RCRS/2024/RCRS2024_0897.pdf}} developed for the SPOTLIGHT\footnote{SPOTLIGHT survey webpage: \url{https://spotlight.ncra.tifr.res.in/}} project, we simulated several types of uGMRT beams and their size variation as a function of elevation for uGMRT Band-3 and Band-4 on NGC~6093 as a reference GC.

Based on the beam simulation results keeping a balance between sensitivity and sky coverage the uGMRT Band-3 and Band-4 central square's (CSQ) phased array (PA) beams were the most suitable bands and primary beam type for designing the survey. A PA beam is formed by adding all the antenna voltages coherently preserving the phase information, making the beam narrower but more sensitive than an incoherent array (IA) beam (PA beam sensitivity is $\sqrt{N}$ times IA beam sensitivity, where $N$ is the number of antennae used to form the beam) that is formed by incoherently adding all the antenna power without the phase information. A CSQ PA beam is a PA beam that is formed only using the closely packed (within 1 km$^2$) 14 uGMRT central square (CSQ) antennas, which are wider than the PA beams formed by adding the long-spaced arm antennas.

\begin{figure}
    \includegraphics[width=\columnwidth]{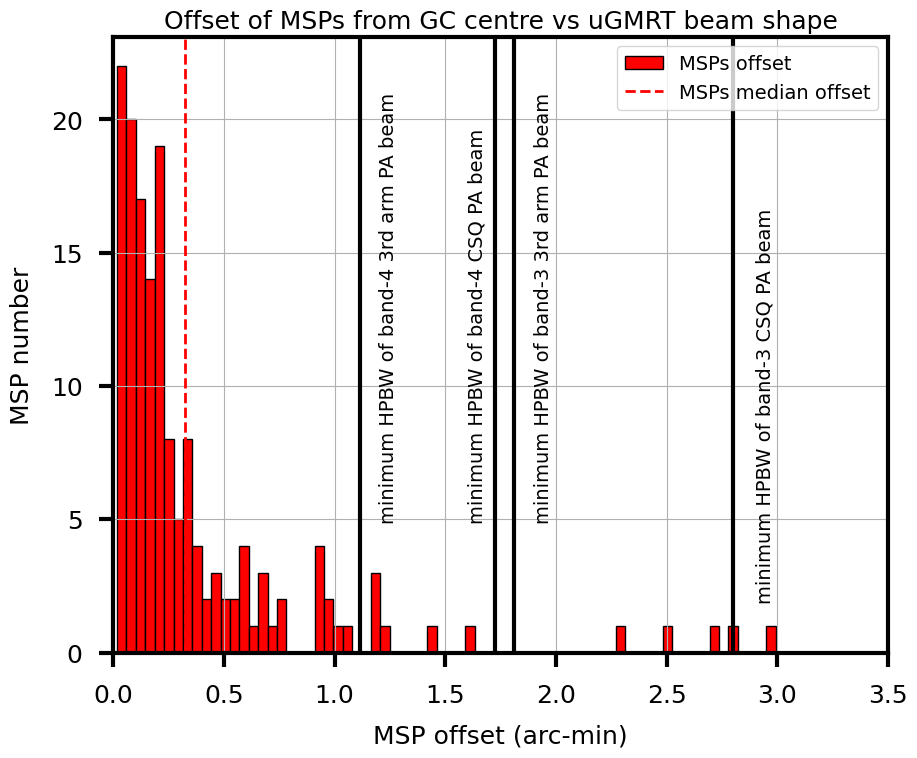}
    \caption{Histogram of the reported position offsets of all GC MSPs from the cluster centre. The back solid lines are the same as Figure \ref{fig:GCradiusvsbeamsize}}
    \label{fig:MSPoffsetvsbeamsize}   
    \includegraphics[width=\columnwidth]{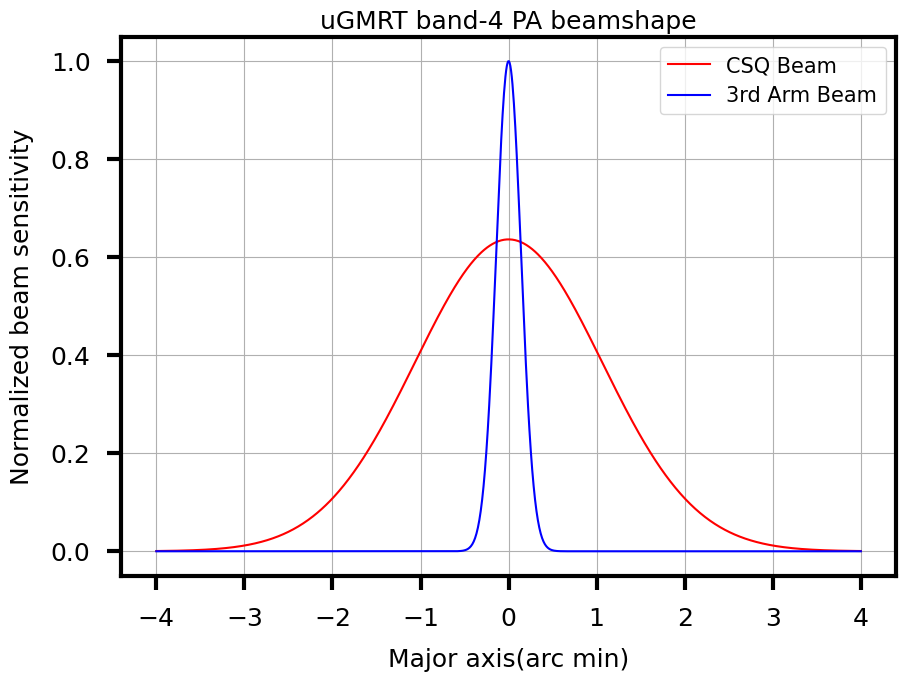}
    \caption{The uGMRT Band-4 CSQ and 3rd arm PA beam's major axis (calculated using {\tt uGMRT\_beam\_fov.py} for the same initial condition) versus normalised beam sensitivity (to 3rd arm PA beam sensitivity), where it is assumed that the beam sensitivity response is a Gaussian response.}
    \label{fig:CSQvsthirdArm}
\end{figure}
Statistically, as illustrated in Figure \ref{fig:GCradiusvsbeamsize}, the minimum Half Power Bandwidth (HPBW) of the Band-4 PA beam exceeds the half-light radius for 77\% of GCs, while for Band-3, it does so for 92\%. We expect to find our MSPs close to the core of the GC. In the observed MSP distribution (as shown in Figure \ref{fig:MSPoffsetvsbeamsize}), 96\% of all the reported localized GC MSPs have a positional offset\footnote{All the MSP position offset data taken from Paulo Freire GC website, see: \url{https://www3.mpifr-bonn.mpg.de/staff/pfreire/GCpsr.html}} from the GC centre that is less than the minimum HPBW of uGMRT CSQ Band-4 PA beam and 86\% for them have offset value less than HPBW/2 (indicating they are residing in the GC core region). This result is not a selection effect: in most surveys - especially the Parkes and GBT surveys, and the Arecibo searches at lower frequencies - the search beams' HPBWs are larger than 4 arcminutes. For Band-3, this coverage is even better, statistically establishing that to observe the whole target GC with optimized sensitivity, Band-3 or Band-4 CSQ PA beam is sufficient.

Between Band-3 and Band-4, we preferred Band-4 as the first choice due to its lower system temperature (for 650 MHz, $T_\mathrm{sys}\, \sim\, 102.5 \rm\, K$) compared to Band-3 (for 400 MHz, $T_\mathrm{sys}\, \sim\, 130 \rm\, K$, reference \cite{Gautam2022}). Additionally, Band-4 experiences significantly less Radio Frequency Interference (RFI), making it the cleanest Band of the uGMRT, whereas Band-3 is more adversely affected by RFI. We use Band-3 only when the target is significantly wider, as it provides better sensitivity coverage compared to Band-4. We excluded Band-5 due to its reduced gain, narrower beam size compared to Band-3 and Band-4 and its focus on high-frequency coverage (where, as discussed in Section \ref{sec:motivation}, low-frequency coverage is preferred).

Along with covering the whole GC with optimum sensitivity, to observe the core with even better sensitivity (where we expect most of our MSPs will be, as we see in Figure \ref{fig:MSPoffsetvsbeamsize}), we decided to simultaneously form another PA beam including up to the third antenna in each arm (i.e. $N \,=\,22$) called the 3rd arm PA beam. As we can see in Figure \ref{fig:CSQvsthirdArm}, due to differences in the baseline length and the number of antennas, the sensitivity of the 3rd arm PA beam is higher than that of the CSQ PA beam (theoretically: number of 3rd arm antennas/number of CSQ antennas = 22/14 $\sim$ 1.6 times). However, the CSQ beam exhibits a wider response compared to the 3rd arm beam. As shown in Figure \ref{fig:GCradiusvsbeamsize}, the uGMRT Band-4 (our primary observing Band) CSQ PA beam's minimum HPBW is larger than the half-light radius for approximately 77\% of GCs, enabling full coverage of the GC with optimal sensitivity. Similarly, for the 3rd arm PA beam, around 87\% of GCs have their core radius within the minimum HPBW, providing optimal coverage of the GC core.

An additional advantage of observing a GC simultaneously with two beams (CSQ and 3rd arm) is that, upon detecting a pulsar in the beam data, the SNR ratio between the two beams can immediately provide an estimate of the pulsar's positional offset from the cluster centre, using the known beam response from our simulation code. For a pulsar sitting exactly in the beam centre, as discussed earlier we would expect the 3rd arm beam SNR will be $\sim$ 1.6 times higher than the CSQ beam SNR and this factor will decrease systematically (as shown in Figure \ref{fig:CSQvsthirdArm}) with a systematic increase of the position offset from the beam centre. With an initial offset estimation, follow-up imaging observations can accurately localise the pulsar in the image plane (see Section \ref{subsec:localization}) with minimal effort.

To meet our sensitivity criteria (discussed in Section \ref{subsec:sensitivity}), we decided to spend 2.5 hours of observing time on each target. Before the target scan, the entire array is phased on a calibrator to form the PA beam. Instrumental and ionospheric phase perturbations necessitate intermediate phasing every 45 minutes, resulting in two phasing sessions over the observing duration. These avoid dephasing and maintain the PA beam's sensitivity during the entirety of the target scans.
This intermediate phasing is done by moving the whole array to a phase calibrator and performing a phasing operation while keeping the beam data recording on. This results in a continuous 2.5 hours beam dataset, with 5$–$10 minutes of phase calibration time every 45 minutes accounting for slewing to the phase calibrator, performing the calibration, and returning to the GCs, which are masked during the analysis.

\subsection{Observation strategy and sensitivity coverage} \label{subsec:sensitivity}
\begin{figure}
\centering
    \includegraphics[width=\columnwidth]{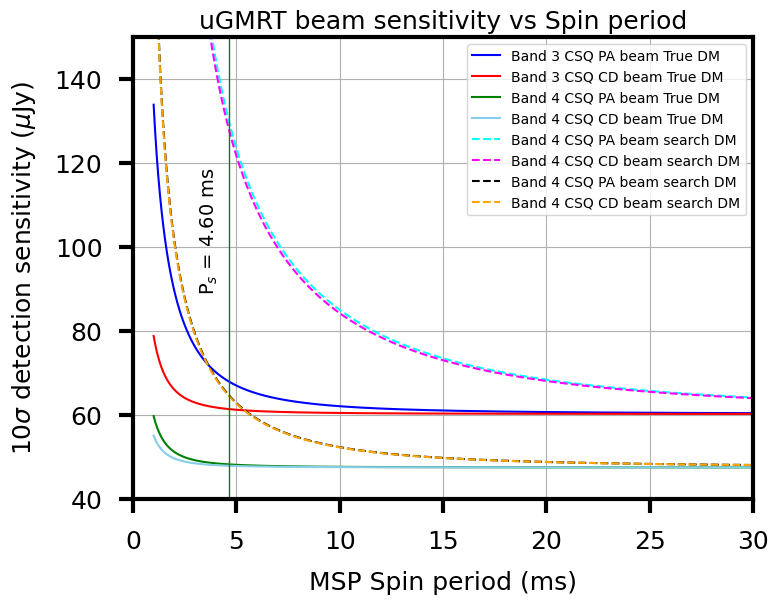}
    \caption{The uGMRT 10$\sigma$ detection sensitivity (in $\mu \rm\, Jy$) versus the spin period of the observed pulsar for both uGMRT bands' CD and PA mode for CSQ beam only. For each observing mode, the true DM sensitivity represents the case when one search DM value corresponds exactly to the pulsar's DM ($\rm DM^*$, which in the calculations above is $\rm 50\:pc\:cm^{-3}$). The search DM sensitivity corresponds to the case when the search DM is $\rm DM^* + DM\:step/2$, where we will have maximum sensitivity loss due to dispersion smearing caused by the finite DM step used in the search. For uGMRT Band-3 and Band-4, the DM steps are 0.05 and 0.1 $\rm pc\:cm^{-3}$ respectively. The sensitivity is calculated using the radiometer equation (refer to Equation \ref{eqn:radiometer}) for a pulsar with 10\% duty cycle. The black vertical line represents the sensitivity achieved for the pulsar with the median spin period of all the GC MSPs, where in Band-3 the DM step smearing introduces a maximum reduction of theoretical sensitivity by a factor of $\sim$ 2, this factor is $\sim$ 1.3 for Band 4.}
    \label{fig:periodvsbeamnoise}
\end{figure}
In GCGPS, we primarily used uGMRT Band-4 (550$-$750 MHz) and for a few wider targets Band-3 (300$-$500 MHz) as discussed in Section \ref{subsec:design}. For a specific Band, the observation strategies employed to achieve optimal target sensitivity are discussed along with the theoretical sensitivity coverage. Given that we are searching for MSPs at mid and low frequencies (with a median GC MSP spin period of 4.60 $\rm ms$\footnote{Data taken from: \url{https://www3.mpifr-bonn.mpg.de/staff/pfreire/GCpsr.html}}), the sensitivity loss due to pulse broadening from scattering and intra-channel dispersion becomes significant.

According to \cite{Bhat2004}, the scatter broadening $W_{scatt}$ in ms is defined as,
\begin{eqnarray} \label{eqn:scattering}
    \log[W_\mathrm{scatt}] & = & -6.46 + 0.154 \log[DM] \nonumber \\
                           &   & + 1.07 \log[DM]^2 - 3.86 \log[\nu],
\end{eqnarray}
Where DM is the dispersion measure in $\rm pc\,cm^{-3}$ and $\nu$ is the observing frequency in $\rm GHz$.

According to \cite{Lorimer2004}, the smearing ($W_\mathrm{DMintra}$) due to intra-channel dispersion can be defined as (when $\Delta\nu << \nu$),
\begin{equation} \label{eqn:DM_intra_smearing}
    W_\mathrm{DMintra} \simeq 8.3 \times DM \times \frac{\Delta\nu}{\nu^3}\: \mu \rm s,
\end{equation}
here $\Delta\nu$ is the channel width/resolution in $\rm MHz$ and $\nu$ is the observing frequency in $\rm GHz$.

While scattering is unavoidable in low-frequency observations, intra-channel dispersion can be eliminated by recording the data in coherently de-dispersed (CD) mode (see \cite{Hankins_1971}), provided the GC's DM is precisely known beforehand. By mitigating intra-channel dispersion smearing, this mode enables the recording of high-time-resolution data by reducing the frequency channels in the filterbank output. 

The GCs targeted in GCGPS primarily lack known pulsars (as discussed in Section \ref{sec:motivation}), with only a few containing a minimal number of MSPs (discussed in Section \ref{subsec:target}). For GCs with precisely known DM (have discovered MSPs), we observe it with two (CSQ and 3rd arm) simultaneously recorded CD beams (coherently dedispersed at the GC MSPs' median DM) having 512 channels and 40.96 $\mu \rm s$ sampling time. For GCs with no known DM, we record two (CSQ and 3rd arm) PA beams with 4096 channels and 81.92 $\mu \rm s$ sampling time.

For each observing mode, the theoretical sensitivity limit can be calculated using the radiometer equation \citep{Dewey_1985}. The minimum detectable flux ($S_\mathrm{min}$) is defined as:
\begin{equation} \label{eqn:radiometer}
    S_\mathrm{min} = \frac{{SNR} \, T_\mathrm{sys} \, \beta}{G N\sqrt{n_\mathrm{pol} BW_\mathrm{eff} \Delta t_\mathrm{obs}}}\sqrt{\frac{W_\mathrm{eff}}{P_\mathrm{s} - W_\mathrm{eff}}}
\end{equation}
We define $S_{min}$ as 10$\sigma$ (i.e. SNR $=$ 10) detection flux. For 8-bit recording, we assume the digitization loss ($\beta$) to be $\sim$ 1. For Band-3 at 400 MHz, $T_\mathrm{sys} \, \sim 130 \rm \, K$,  where for Band-4 at 650 MHz,  $T_\mathrm{sys} \, \sim \, 102.5\rm\, K$ as mentioned earlier in Section \ref{subsec:design}. For uGMRT, the gain per antenna (G) is about 0.35 $\rm \, KJy^{-1}$. As discussed above, $N = 14$ for CSQ and 22 for the 3rd arm PA beams. For total intensity Stokes-I data, the number of polarizations ($n_\mathrm{pol}$) summed up is 2. Our on-source time ($\Delta t_\mathrm{obs}$) is 2.5 hours for each beamforming observation per target. For Band-3 and Band-4, after eliminating the periodic narrow-band RFI-affected channels, the effective Bandwidth ($BW_\mathrm{eff}$) is about 150 $\rm MHz$ and 180  $\rm MHz$ respectively.

Now the effective pulse width can be defined as,
\begin{equation} \label{eqn:effcective_width}
    W_\mathrm{eff} = \sqrt{W_\mathrm{int}^2 + W_\mathrm{DMintra}^2 + W_\mathrm{scatt}^2 + t_\mathrm{samp}^2 + W_\mathrm{DMstep}^2}
\end{equation}
Where $W_\mathrm{int}$ is the intrinsic pulse width, $W_\mathrm{scatt}$ and $W_\mathrm{DMintra}$ (defined in Equation \ref{eqn:scattering} and \ref{eqn:DM_intra_smearing}) are the intra-channel dispersion smearing and scatter broadening time-scale respectively. $t_\mathrm{samp}$ is the smearing that rises due to finite sampling time. $W_\mathrm{DMstep}$ is the smearing that arises due to searching the pulsar with a finite DM step, this is calculated for the worst case scenario, which happens when the pulsar's DM is offset from the search DM by (DM step)/2.

For CD recording $W_\mathrm{DMintra} = 0$ (if the pulsar is exactly at the DM used to dedisperse the data coherently) and $t_\mathrm{samp} = 40.96\:\mu \rm s$ whereas for PA data $W_\mathrm{DMintra}$ have a finite value according to Equation \ref{eqn:DM_intra_smearing}. For both PA and CD, $W_\mathrm{scatt}$ is the same and calculated using Equation \ref{eqn:scattering} at the centre of the band for uGMRT Band-3 and Band-4, which are 0.40 $\rm GHz$, and 0.65 $\rm GHz$ respectively.

For $W_\mathrm{DMstep}$, both PA and CD will suffer the same smearing. According to \cite{Lorimer2004}, for our search, the maximum smearing ($W_\mathrm{DMstep}$) due to the finite DM step can be defined as,

\begin{equation} \label{eqn:DM_step_smearing}
    W_\mathrm{DMstep} = \left(8.3 \times 10^6 \right)\times \left( \frac{DM\:step}{2} \right) \left(\frac{1}{f^2_\mathrm{1}} - \frac{1}{f^2_\mathrm{2} }\right) \: ms,
\end{equation}
Where, $f_\mathrm{1}$, $f_\mathrm{2}$ are the lowest and highest frequencies of the respective band in MHz, which are 300 MHz, 500 MHz for uGMRT Band-3 and 550 MHz, 750 MHz for uGMRT Band-4, respectively. We take the DM step to be 0.05 $\rm pc\:cm^{-3}$ for Band-3 and 0.1 $\rm pc\:cm^{-3}$ for Band-4 according to our DM plan (which is the maximum DM step for the respective band, see Section \ref{sec:dataanalysis}).
 
\begin{table*}
\vspace{-10pt}
\centering
\caption{Sensitivity estimation of GCGPS survey using the Radiometer equation by \cite{Dewey_1985}, for all the GCGPS observing modes and bands with typical on-source time per target. The notations BM1 and BM2 represent two beams used under this survey (for reference see Table \ref{table:targets}). The presented estimated sensitivity is calculated for the median spin period of all the GC MSPs (median P$_\mathrm{s}$ = 4.60 $\rm ms$), for a DM value of 50 $\rm pc\,cm^{-3}$ with a duty cycle of 10\%. S$_{min}$ represents the 10$\sigma$ minimum detection sensitivity for the pulsar detected in the search at the true pulsar DM.}
\begin{tabular}{|c|c|c|c|c|c|c|c|}
\toprule
\midrule
On-source & Spin period & uGMRT Band & uGMRT array & Beam type & Sampling & Channel & S$_{min}$ (10$\sigma$)\\
time (hours) & (ms) & (Band-3/Band-4) & (BM1/BM2) & (PA/CD) & ($\mu$s) & (number) & ($\mu$ Jy) \\
\midrule
\multirow{8}{*}{2.5} & \multirow{8}{*}{4.60} & \multirow{4}{*}{Band-3} & \multirow{2}{*}{BM1 (CSQ)} & PA & 81.92 & 4096 & 68 \\
\cline{5-8}
    & & & & CD & 40.96 & 512 & 61 \\
\cline{4-8}
    & & & \multirow{2}{*}{BM2 (3rd Arm)} & PA & 81.92 & 4096 & 43 \\
\cline{5-8}
    & & & & CD & 40.96 & 512 & 39 \\
\cline{3-8}
    & & \multirow{4}{*}{Band-4} & \multirow{2}{*}{BM1 (CSQ)} & PA & 81.92 & 4096 & 48 \\
\cline{5-8}
    & & & & CD & 40.96 & 512 & 47 \\
\cline{4-8}
    & & & \multirow{2}{*}{BM2 (3rd Arm)} & PA & 81.92 & 4096 & 30 \\
\cline{5-8}
    & & & & CD & 40.96 & 512 & 30 \\
\bottomrule
\end{tabular}
\label{table:sesitivity}
\end{table*}

For sensitivity analysis, we use intrinsic pulse width ($W_\mathrm{int}$) as 10\% of the pulse period ($P_\mathrm{s}$) and DM = 50 $\rm pc \, cm^{-3}$, which is a reasonable estimation for a low-frequency MSP survey like the GCGPS. Figure \ref{fig:periodvsbeamnoise} shows the detection sensitivity for both PA and CD modes for CSQ beams only, at true pulsar DM (here DM = 50 $\rm pc\:cm^{-3}$) as well as detection with the maximum DM offset of ${\rm DM\:step/2}$, for both uGMRT bands, as a function of pulsar spin period (P$_\mathrm{s}$). For the 3rd arm beam, for every observing mode, these sensitivities will scale down (i.e. more sensitive) by a factor of $\sim$1.6 (as it scales as $N_2/N_1 = 22/14$, see Section \ref{subsec:design}).

We present the theoretical 10$\sigma$ detection flux for the GCGPS survey for different observation modes for a 4.60 $\rm ms$ MSP (median spin period of all GC pulsars) with a DM value of 50 $\rm pc\,cm^{-3}$ for both uGMRT bands when the pulsar detected at it's true DM, in Table \ref{table:sesitivity}. Whereas, the maximum sensitivity loss for the search with finite DM step, for Band-3 (DM step = 0.05 $\rm pc\: cm^{-3}$) is around $\sim$ 2 times and for Band-4 (DM step = 0.1 $\rm pc\: cm^{-3}$) its $\sim$ 1.3 times (for the same 4.60 $\rm ms$ candidate period) compared to the sensitivity  (detection at the true DM) we quoted in Table \ref{table:sesitivity} (for reference see Figure \ref{fig:periodvsbeamnoise}).

\subsection{Target selection} \label{subsec:target}
As discussed in Section \ref{sec:motivation}, currently the most prominent ongoing GC search surveys are MeerKAT's TRAPUM and FAST's GC survey. According to \cite{Ridolfi2022}, for 2 hours of on-source time, the minimum detectable flux (10$\sigma$) is 14 $\mu \rm Jy$ and 10 $\mu$Jy for MeerKAT's UHF (544$–$1087 MHz) and L-Band (856$–$1711 $\rm MHz$) respectively. Whereas according to \cite{Pan2021a}, for FAST's L-Band(1.0 $\rm GHz$-1.5 $\rm GHz$) survey is 0.76 $\mu \rm Jy$ (same on-source time and detection limit). Assuming the pulsar spectral index $-2$, the TRAPUM's L-Bands (the most used Band) scaled sensitivity at 650 $\rm MHz$ (uGMRT Band-4) is about $\sim$ 39 $\mu \rm Jy$, and for FAST it's about 2.81 $\mu \rm Jy$. Now from \cite{Ridolfi_2022}, for the same on source time (2 hours), duty cycle (8\%), DM (52 $\rm pc\,cm^{-3}$) and assuming the pulsar period about 10 $\rm ms$, the GCGPS primary beam's (Band-4 CSQ PA beam) theoretical 10$\sigma$ sensitivity limit is about 36 $\mu \rm Jy$.

As the FAST GC project follows observing GCs independent of the fact of having discovered MSPs in them or not, along with its superior sensitivity compared to the GCGPS survey, we completely excluded the GCs that can be observed by the FAST (i.e.  $\delta > -17^\circ$) for our survey. Also, due to the comparable sensitivity of MeerKAT's TRAPUM for pulsars having a high spectral index, we exclude the GCs that MeerKAT has already observed. Now the remaining GCs were the GCs visible to the uGMRT within the declination range $-53^\circ < \delta < -17^\circ$ ($\delta > -53^\circ$, is the uGMRT sky coverage to have at least 2 hours of target on-source time), and not observed under the TRAPUM project.

\begin{figure}
\centering
    \includegraphics[width=\columnwidth]{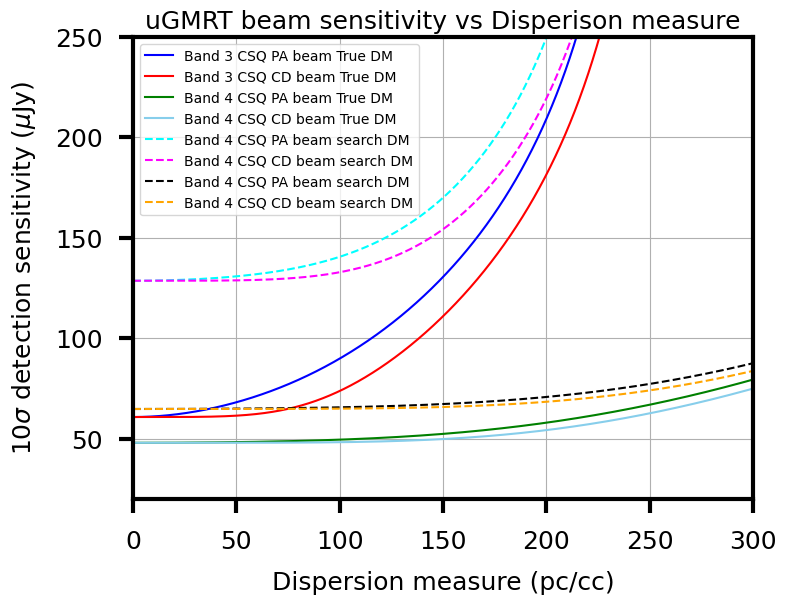}
    \caption{Plot representing the uGMRT 10$\sigma$ detection sensitivity versus DM variation for a fixed pulsar period and duty cycle of 4.60 $\rm ms$ (median spin period of all GC MSPs) and 10\% respectively, for PA and CD data for both uGMRT bands for CSQ beams only in different colours. The sensitivities at true and search DM differ in a similar way as seen in Figure \ref{fig:periodvsbeamnoise}. Band-3 CSQ beam is the most affected band with  high DM, and DM smearing (due to finite search DM step) as well, where Band-4 is comparatively better in high DMs as well.}
    \label{fig:DMvsbeamnoise}
\end{figure}
Most of the remaining GCs had no prior reported MSP search observations, while a subset had been previously observed with the GBT's 2 $\rm GHz$ survey or Parkes' 1.4 $\rm GHz$ GC survey. Assuming an MSP spectral index of $-2$ (with GBT 2 $\rm GHz$ and Parkes 1.4 $\rm GHz$ sensitivities scaled to uGMRT Band-4 at 650 $\rm GHz$), for the same on-source time the theoretical sensitivity of GCGPS is approximately 3 times better than that of GBT and about 8 times better than the Parkes survey. Therefore, we considered observations of the GBT and Parkes observed GCs in the GCGPS (uGMRT Band-3 and Band-4) frequencies.

Figure \ref{fig:DMvsbeamnoise} illustrates significant sensitivity loss in both Band-3 and Band-4, primarily for DM smearing due to the finite search DM step, along with intra-channel dispersion and scattering at higher DMs. However, Band-4 maintains moderate sensitivity even at higher DMs, which was one of the key reasons for selecting it as our primary observing band over Band-3. To account for this sensitivity loss, from the remaining GCs, we chose targets with known or predicted maximum DMs of $\sim$ 250 $\rm pc\,cm^{-3}$  for Band-4. For broader targets requiring Band-3 observations, the maximum DM (known or predicted) was limited to less than 100 $\rm pc\,cm^{-3}$. We also note that the scatter-broadening timescale predicted by Equation \ref{eqn:scattering} does not account for the possibility that GCs, being located at higher Galactic latitudes, may experience reduced scattering even for distant sources with higher DMs. This results in shorter scattering timescales observed for GC pulsars compared to field pulsars with equivalent DMs \citep{He_2024}. This will result in sensitivity loss even lower than the value we got using Equation \ref{eqn:scattering}, \ref{eqn:DM_intra_smearing}, \ref{eqn:radiometer}, \ref{eqn:effcective_width}, and \ref{eqn:DM_step_smearing}.

Finally, as the MSPs can be a prominent source of X-ray/gamma-ray emissions \citep{Freire_2011, Johnson_2013, Zhang_2003, Zhang_2022}, we ranked the GCs based on reported X-ray/gamma-ray brightness from different high-energy surveys \citep{Webb_2006, Adbo_2010, Tam_2011, Tam_2016}. GCs without any discovered pulsars but exhibiting very high gamma-ray brightness are ideal targets for our survey. Compactness and higher total mass were also considered to prioritise the GC observations. We also assigned lower priority to GCs located more than $\sim$ 15 kiloparsecs ($\rm kpc$) away or with a half-light radius exceeding twice the HPBW of the uGMRT CSQ PA beam. After applying these criteria, 31 GCs were identified as optimal candidates for the GCGPS survey. In Table \ref{table:targets}, we present these 31 GCs with their relevant information, which will be observed (over multiple uGMRT cycles) under GCGPS. In Table \ref{table:targets}, we also present the expected observing mode for each GC, which was decided based on the discussion presented in Section \ref{subsec:sensitivity}.

\begin{table*}
\vspace{-10pt}
\centering
\caption{List of target globular clusters of the GCGPS survey based on the selection criteria listed in the text. All relevant values are taken from the W. E. Harris list of GC parameters, \citep[2010 edition]{Harris_1996}. $R_\mathrm{SUN}$ is the GC distance from the Sun. The DM values in the second-last column of the table either represent the median DM of all the MSPs in that particular GC (GCs with already known MSPs), or for GCs with no known pulsars the DM predicted by the YMW16 model \citep{Yao_2017} (marked with \enquote{*}). The last column represents the expected observing mode for each GC, which we decided based on the GC parameters, where PA and CD are the beam type, BM1 and BM2 are the CSQ and the 3rd Arm beams, and the two numbers (3, and 4) represents the two uGMRT observing bands (Band-3, and Band-4) respectively.}
\begin{tabular}{|c|c|c|c|c|c|c|c|}
\toprule
\midrule
GC name & RA (J2000) & Dec (J2000) & Core radius & Half-light radius & Distance  & GC DM & Expected observing \\
(NGC XXXX) & (hh mm ss) & (dd mm ss) & (arc-min) & (arc-min) & $R_\mathrm{SUN}$ (kpc) & (pc cm$^{-3}$) & mode\\
\midrule
NGC~1904 & 05 24 11.09 & -24 31 29.0 & 0.16 & 0.65 & 12.9 & 53.1* & PA (BM1, BM2)-4\\
\midrule
NGC~4590 & 12 39 27.98 & -26 44 38.6 & 0.58 & 1.51 & 10.3 & 50.3* & PA (BM1, BM2)-4\\
\midrule
NGC~5286 & 13 46 26.81 & -51 22 27.3 & 0.28 & 0.73 & 11.7 & 146.3* & PA (BM1, BM2)-4\\
\midrule
NGC~5897 & 15 17 24.50 & -21 00 37.0 & 1.40 & 2.06 & 12.5 & 48.6* & PA (BM1, BM2)-3\\
\midrule
NGC~5986 & 15 46 03.00 & -37 47 11.1 & 0.47 & 0.98 & 10.4 & 92.17 & CD (BM1, BM2)-4\\
\midrule
NGC~6093 & 16 17 02.41 & -22 58 33.9 & 0.15 & 0.61 & 10.0 & 75.2* & PA (BM1, BM2)-4\\
\midrule
NGC~6121 & 16 23 35.22 & -26 31 32.7 & 1.16 & 4.33 & 2.2 & 62.86 & CD (BM1, BM2)-3\\
\midrule
NGC~6139 & 16 27 40.37 & -38 50 55.5 & 0.15 & 0.85 & 10.1 & 166.6* & PA (BM1, BM2)-4\\
\midrule
NGC~6273 & 17 02 37.80 & -26 16 04.7 & 0.43 & 1.32 & 8.8 & 125.9* & PA (BM1, BM2)-4\\
\midrule
NGC~6284 & 17 04 28.51 & -24 45 53.5 & 0.07 & 0.66 & 15.3 & 142.5* & PA (BM1, BM2)-4\\
\midrule
NGC~6287 & 17 05 09.13 & -22 42 30.1 & 0.29 & 0.74 & 9.4 & 115.4* & PA (BM1, BM2)-4\\
\midrule
NGC~6293 & 17 10 10.20 & -26 34 55.5 & 0.05 & 0.89 & 9.5 & 146.4* & PA (BM1, BM2)-4\\
\midrule
NGC~6304 & 17 14 32.25 & -29 27 43.3 & 0.21 & 1.42 & 5.9 & 153.3* & PA (BM1, BM2)-4\\
\midrule
NGC~6316 & 17 16 37.30 & -28 08 24.4 & 0.17 & 0.65 & 10.4 & 186.1* & PA (BM1, BM2)-4\\
\midrule
NGC~6333 & 17 19 11.26 & -18 30 57.4 & 0.45 & 0.96 & 7.9 & 114.2* & PA (BM1, BM2)-4\\
\midrule
NGC~6342 & 17 21 10.08 & -19 35 14.7 & 0.05 & 0.73 & 8.5 & 122.4* & PA (BM1, BM2)-4\\
\midrule
NGC~6355 & 17 23 58.59 & -26 21 12.3 & 0.05 & 0.88 & 9.2 & 182.6* & PA (BM1, BM2)-4\\
\midrule
NGC~6356 & 17 23 34.93 & -17 48 46.9 & 0.24 & 0.81 & 15.1 & 144.4* & PA (BM1, BM2)-4\\
\midrule
NGC~6388 & 17 36 17.23 & -44 44 07.8 & 0.12 & 0.52 & 9.9 & 171.4* & PA (BM1, BM2)-4\\
\midrule
NGC~6528 & 18 04 49.64 & -30 03 22.6 & 0.13 & 0.38 & 7.9 & 206.9* & PA (BM1, BM2)-4\\
\midrule
NGC~6541 & 18 08 02.36 & -43 42 53.6 & 0.18 & 1.06 & 7.5 & 108.4* & PA (BM1, BM2)-4\\
\midrule
NGC~6553 & 18 09 17.60 & -25 54 31.3  & 0.53 & 1.03 & 6.0 & 227.3* & PA (BM1, BM2)-4\\
\midrule
NGC~6569 & 18 13 38.80 & -31 49 36.8 & 0.35 & 0.80 & 10.9 & 175.7* & PA (BM1, BM2)-4\\
\midrule
NGC~6637 & 18 31 23.10 & -32 20 53.1 & 0.33 & 0.84 & 8.8 & 119.8* & PA (BM1, BM2)-4\\
\midrule
NGC~6638 & 18 30 56.10 & -25 29 50.9 & 0.22 & 0.51 & 9.4 & 155.5* & PA (BM1, BM2)-4\\
\midrule
NGC~6642 & 18 31 54.10 & -23 28 30.7 & 0.10 & 0.73 & 8.1 & 155.8* & PA (BM1, BM2)-4\\
\midrule
NGC~6652 & 18 35 45.63 & -32 59 26.6 & 0.10 & 0.48 & 10.0 & 116.6* & PA (BM1, BM2)-4\\
\midrule
NGC~6681 & 18 43 12.76 & -32 17 31.6 & 0.03 & 0.71 & 9.0 & 105.2* & PA (BM1, BM2)-4\\
\midrule
NGC~6717 & 18 55 06.04 & -22 42 05.3 & 0.08 & 0.68 & 7.1 & 103.3* & PA (BM1, BM2)-4\\
\midrule
NGC~6723 & 18 59 33.15 & -36 37 56.1 & 0.83 & 1.53 & 8.7 & 82.0* & PA (BM1, BM2)-4\\
\midrule
NGC~6809 & 19 39 59.71 & -30 57 53.1 & 1.80 & 2.83 & 5.4 &  56.7* & PA (BM1, BM2)-3\\
\bottomrule
\end{tabular}
\label{table:targets}
\end{table*}
We note that we are planning to include the GCs observed with MeerKAT in our survey in the near future. This will be facilitated by the improved sensitivity and sky coverage offered by the multi-beam formation using the full uGMRT array, utilising all 30 antennas instead of the single-beam formation by the 14 antennae, which is currently in use in GCGPS. The key enabler of this enhancement is the SPOTLIGHT backend\footnote{For relevant information about the SPOTLIGHT project, see \url{https://spotlight.ncra.tifr.res.in/}}, which allows target-specific multi-beam formation with a maximum of 200 post-correlation beams\footnote{See \url{https://www.ursi.org/proceedings/RCRS/2024/RCRS2024_0803.pdf}}\footnote{See \url{https://www.ursi.org/proceedings/RCRS/2024/RCRS2024_1101.pdf}}. Additionally, we will deploy the in-field phasing technique developed by \cite{Kudale_2024}, ensuring that the highly sensitive PA beam (formed using all 30 antennas) maintains its sensitivity over long on-source time. These advancements will enhance our survey sensitivity by a factor of 2$-$3 for uGMRT Band-3 and Band-4, compared to the current sensitivity of the GCGPS.

This will enable us to observe all low-DM GCs (DM $<$ 250 $\rm pc, cm^{-3}$) seamlessly within the declination range $-53^\circ < \delta < -17^\circ$.

In this paper, we present the discovery of the first MSP in NGC~6093 from the GCGPS, along with its timing studies. A comprehensive list of already GCGPS observed GCs from Table \ref{table:targets}, including all discoveries and non-detection limits, will be detailed in future publications.

\section{Data Analysis} \label{sec:dataanalysis}
The PA beam data recorded from the uGMRT GWB backend \citep{Reddy_2017} was initially processed to mitigate RFI signals and to correct instrumental effects, such as slopes in the bandpass. We used {\tt GPTOOL}\footnote{See \url{https://github.com/chowdhuryaditya/gptool}} to eliminate RFI signals exceeding the threshold value along both the time and frequency axes, as well as to normalise the Band shape. The Bandpass-corrected filtered beam data was then converted to the {\tt SIGPROC}\footnote{See \url{https://sigproc.sourceforge.net/}} filterbank format for further processing. We developed an end-to-end {\tt  PRESTO}\footnote{PRESTO GitHub page: \url{https://github.com/scottransom/presto}}-based pulsar search pipeline called {\tt PulsarSearchScript} ($\rm PSS$), which is available on GitHub\footnote{See \url{https://github.com/jyotirmoydas5392/Pulsar_Search_Script}}. This pipeline is integrated with a candidate sifting code, enabling the reduction of the huge number of candidates (based on some thresholds and grouping criteria discussed below) to search for pulsars efficiently. We used the pulsars discovered by the GHRSS survey \citep{Bhattacharyya_2016, Bhattacharyya_2019} to test and optimise the PSS pipeline used in GCGPS.

For the observed GCs that already had discovered pulsars, for both uGMRT bands (Bnad-3, and BNad-4), we searched within a range of the median DM $\pm$ 15$\,\rm pc\,cm^{-3}$ with a DM step of 0.02$\,\rm pc\,cm^{-3}$, resulting in a total of 1500 DM trials, while having negligible additional sensitivity loss due to search with finite DM step size, and no-intrachannel dispersion sensitivity loss also due to CD data.

After de-dispersion, we performed a Fast Fourier Transform (FFT) on the de-dispersed data, followed by an acceleration search with a $\pm$ 200(z$_\mathrm{max}$) bin drift correction to account for the Doppler-shifted period caused by the line-of-sight (LOS) acceleration of a binary system.

After the acceleration search, to reduce the enormous number of candidates, we apply a DM clustering filter in our candidate-sifting code. To optimise our sifting code, we defined the DM clustering threshold for each band based on a worst-case scenario: a candidate with a 1 ms period and a 5$\sigma$ detection significance, representing the lower limits of our search parameter space. Based on tests using true pulsar data from the GHRSS survey, we adopted a conservative assumption that such signals remain detectable within $\pm\: 0.05\:\mathrm{pc\:cm^{-3}}$ for Band-3 and $\pm\:0.15\:\mathrm{pc\:cm^{-3}}$ for Band-4 of their true DM values. Beyond these ranges, dispersion smearing causes the signal to become undetectable in {\tt PRESTO}'s acceleration search. This sets the total DM span for clustering to $0.1\: \mathrm{pc\:cm^{-3}}$ for Band-3 and $0.3\: \mathrm{pc\:cm^{-3}}$ for Band-4. For both uGMRT bands, these spans define the minimum DM space a real pulsar should occupy with a period of 1 ms and a detection significance of 5$\sigma$. Candidates with longer periods or higher SNRs are less affected by smearing and are expected to meet these thresholds easily. For GCs with known MSPs (i.e., known DMs), the clustering threshold for uGMRT Band-4 is set to 15, based on a DM step size of $0.02\:\rm pc,cm^{-3}$ and a total clustering span of $0.3\:\rm pc\:cm^{-3}$ (i.e., 0.3/0.02). So, for Band-4, for the DM step 0.02 $\rm pc\:cm^{-3}$, a candidate has to be present in at least 15 consecutive DM trials to qualify for further analysis. Similarly, for Band-3, we follow the same approach using the defined band-specific DM step size and DM span. The upper periodicity limit was set to 10 seconds.

After this sifting, we were left with a manageable number of candidates, yet making sure we didn't miss any true pulsar candidates. Now we fold these candidates over frequency and time domains using {\tt PRESTO}'s {\tt prepfold}. Then, manual inspections were conducted over these folded profiles to check for the presence of any periodic signals in both time and frequency domains.

To process the data from GCs without any discovered pulsars, we initially used both the NE2001 \citep{Cordes_2001} and YMW16 \citep{Yao_2017} models to predict the GC DM and optimise our DM search range.
Testing both models on GCs with known DMs revealed that both tend to overpredict the GC DM at the given GC distance. However, in most cases, the YMW16 model provided DM predictions closer to the actual values compared to the NE2001 model. In the worst-case scenario, the YMW16 model predicted DM was a factor of 1.5$-$2 higher than the GC DM. This observation led us to use the YMW16 model over the NE2001 to predict the DM values for further analysis.

Using the predicted DM value (DM$^{*}$), we searched over the DM space as given below (optimising computational cost):
\begin{equation}
\label{eqn:DM_plan1}
    \frac{1}{3} DM^{*}  \leq  DM \leq \frac{3}{2} DM^{*},
\end{equation}
or
\begin{equation}
\label{eqn:DM_plan2}
    DM =  DM^{*} \pm 50\, \rm pc \, cm^{-3}
\end{equation}

We chose our DM space according to Equation \ref{eqn:DM_plan1} or Equation \ref{eqn:DM_plan2}, whichever probes a larger DM range. As shown in Equation \ref{eqn:DM_plan1}, for larger predicted DM values, the DM search range is biased towards lower DMs rather than higher DMs due to the overpredictive nature of the YMW16 model. For GCs with no pulsars, we have used a DM step of 0.05$\,\rm pc\,cm^{-3}$, and 0.1 $\rm pc\:cm^{-3}$ for uGMRT Band-3 and Band-4, respectively, optimising the computational costs while keeping a manageable sensitivity loss for smearing due to search with finite DM step size. After dedispersion and acceleration search, we applied the candidate sifting approach as discussed earlier. We processed the data through $\rm PSS$  using this DM plan, applying a 5$\sigma$ candidate detection threshold, the appropriate DM clustering threshold (as outlined earlier in this section), and period cutoffs of 1 ms and 10 seconds. The filtered candidates were then folded and manually inspected for the presence of periodic signals in time and frequency space. These search strategies were systematically applied to all of our GCGPS data.

\begin{figure*}
\centering
    \includegraphics[width=\textwidth]{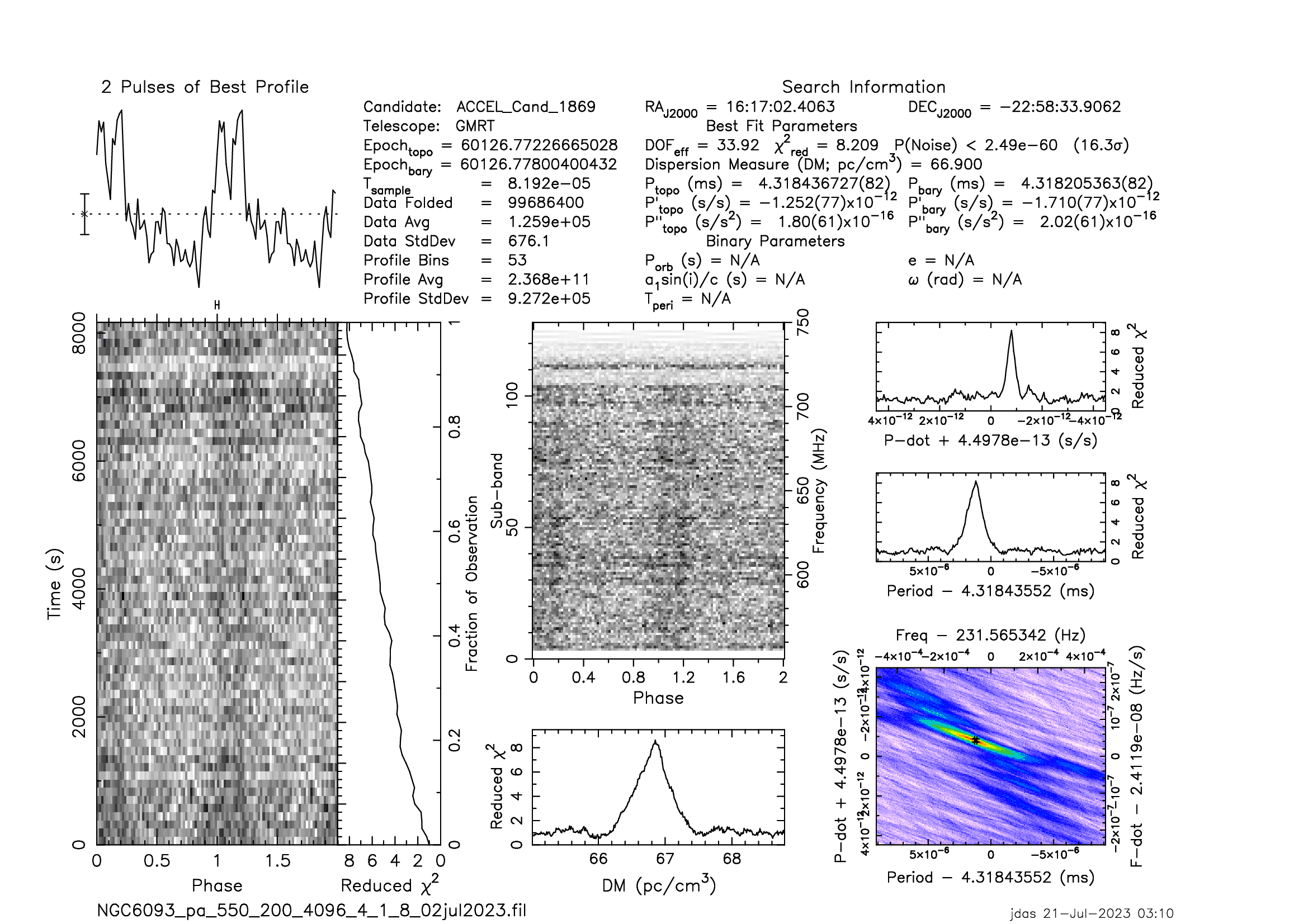}
    \caption{The discovery  {\tt  PRESTO}  output plot of J1617$-$2258A, the first MSP of NGC6093.  The time versus phase and frequency versus phase output are presented in greyscale in the same figure.}
    \label{fig:NewMSPplot}
\end{figure*}

\section{First pulsar detected in NGC~6093 (M80)}
\label{sec:NGC6093}
\subsection{Discovery}
\label{subsection:discovery}
Under GCGPS, in July 2023 we observed NGC~6093, using uGMRT Band-4 CSQ and 3rd arm PA beams simultaneously. At first, we performed the search on the CSQ data having wider beam response, using the YMW16 predicted value for NGC~6093 (DM* $\sim$ 75.2$\,\rm pc\,cm^{-3}$), following Equation \ref{eqn:DM_plan2} over the DM range of 25 to 125$\,\rm pc\,cm^{-3}$, with DM step 0.1$\,\rm pc\,cm^{-3}$. During the manual inspection of the folded profiles in the search output, we identified an MSP at a period of 4.32 ms with a DM of 66.9 $\,\rm pc\,cm^{-3}$. With 2.5 hours of on-source time, the CSQ beam resulted in a detection SNR of 16.3$\sigma$, whereas the 3rd arm beam folding for this MSP (with the known period, period derivative and DM apriori from CSQ detection) yielded a detection significance of 6.3$\sigma$. This difference in SNR suggests a positional offset of the MSP from the cluster centre (details in Section \ref{subsec:localization}). In this region, the sensitivity of the 3rd arm beam decreases rapidly, whereas the wider CSQ beam maintains moderate sensitivity, allowing better coverage.
In Figure \ref{fig:NewMSPplot}, we present the discovery folded profile plot of J1617$-$2258A (the first pulsar of M80, henceforth J1617$-$2258A), from the standard {\tt  PRESTO} folding output.

\subsection{Localization}
\label{subsec:localization}

\begin{figure}
\centering
    \includegraphics[width=\columnwidth]{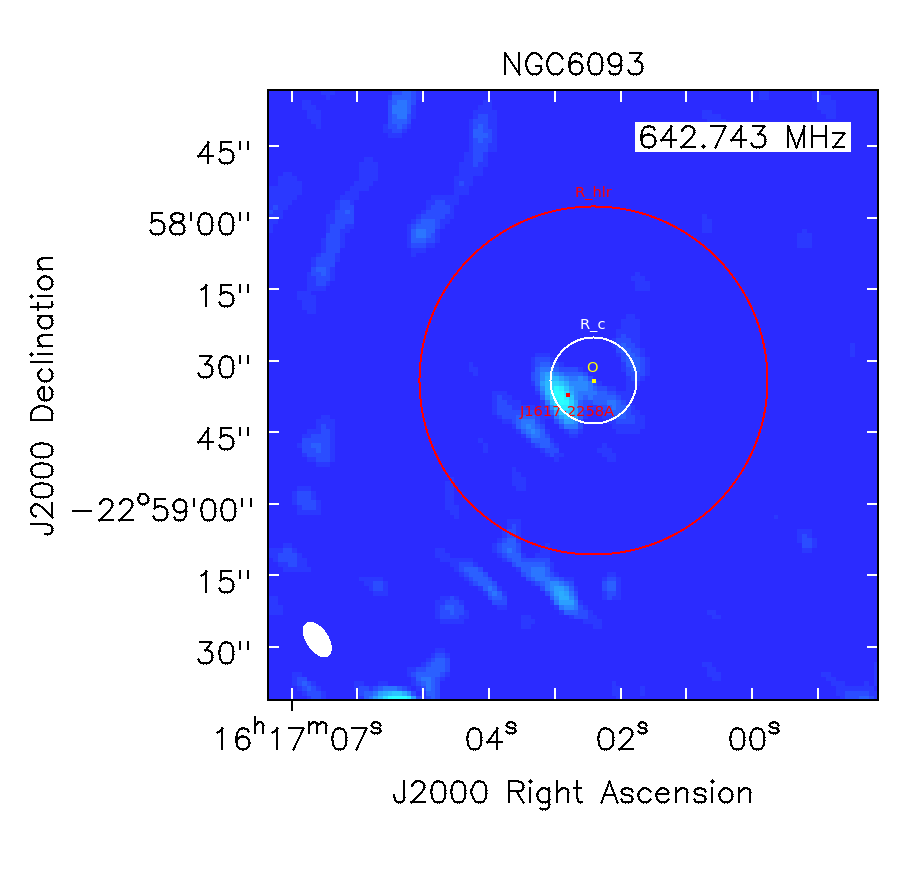}
    \caption{The uGMRT Band-4 ($f_\mathrm{c} = 650\:\rm MHz$) image of M80 using the CAPTURE pipeline. The image was produced from the simultaneous recording of the beam data during the beamforming discovery observation. The yellow and the red circles represent the core radius and the half-light radius of M80 respectively. The point marked with \enquote{O} is the GC centre and the position of J1617$-$2258A is obtained from the timing analysis marked with \enquote{A}, which is sitting inside the core radius on the lower-left side.}
    \label{fig:NewMSPimage}
\end{figure}
In the GCGPS observations, simultaneous to the beam recording we also record the interferometric data with a time integration of 5.3 sec for imaging analysis of the field. The imaging is done using the automated pipeline, called CAPTURE \citep{Kale_2021} developed for the uGMRT data. For the M80 observations, we observed 3C48 as the amplitude calibrator, and 1830$-$360 was observed for every 40 minutes of the target scan (scan of NGC~6093) as a phase calibrator. 

The uGMRT interferometric imaging for 2.2 hours of integration time on M80, resulted in an rms noise of $\sim\,20\,\mu\,\rm Jy$. Imaging of the field revealed 4 sources within a 1.5 arc-min radius from the cluster centre. The first source (the source visible in Figure \ref{fig:NewMSPimage}) was $\sim$ 6.5$''$ off from the cluster centre where the discovery beam was formed. The approximate distance of the source was calculated using the SNR ratio of both the beams (see Section \ref{subsection:discovery}) and the beam simulation code {\tt uGMRT\_beam\_fov.py} (see Section \ref{subsec:design}). We simulated the Band-4 CSQ and the 3rd Arm PA beam for the time of observation to determine the offset from the beam centre. The simulation indicates that the source is located within approximately 10 arc-seconds of the cluster centre. Now the source with a $\sim$ 6.5$''$ off from the cluster centre, and an imaging estimated flux of 350 $\pm$ 25 $\mu\rm Jy$ was considered as the primary target as the position of the pulsar.

To verify this position estimate, we observed NGC 6093 using both the CSQ and 3rd Arm PA beams, centred on the probable pulsar location (i.e. to the source, $\sim$ 6.5$''$ off from the cluster centre). For 1.5 hours of on-source time, the CSQ and 3rd Arm PA beam detection significance was 12.2$\sigma$ and 17.3$\sigma$ respectively (i.e. 3rd arm to CSQ SNR ratio $\sim$ 1.42). Now for the pulsar to be localised, the expected SNR ratio is $\sim$ 1.6 (see Figure \ref{fig:CSQvsthirdArm} and discussion in Section \ref{subsec:design}). We note that the theoretical and observed SNR ratio difference may arise due to multiple factors like the phasing efficiency, available antennas, RFI condition for both beams, etc. Therefore, we identified this radio source as J1617$-$2258A and proceeded with timing follow-up to enhance astrometric precision.

More than one year of timing follow-up of J1617$-$2258A allowed us to obtain a phase-coherent timing solution (discussed in Section \ref{subsec:timing}), improving the positional error(1$\sigma$) up to 0.30$''$. The timing position is marked in Figure \ref{fig:NewMSPimage} with the letter \enquote{A}, which overlaps with the imaging position, confirming that it is J1617$-$2258A.

\subsection{Scattering timescale and profile evolution}
\label{subsec:profile}

\begin{figure}
\centering
    \includegraphics[width=\columnwidth]{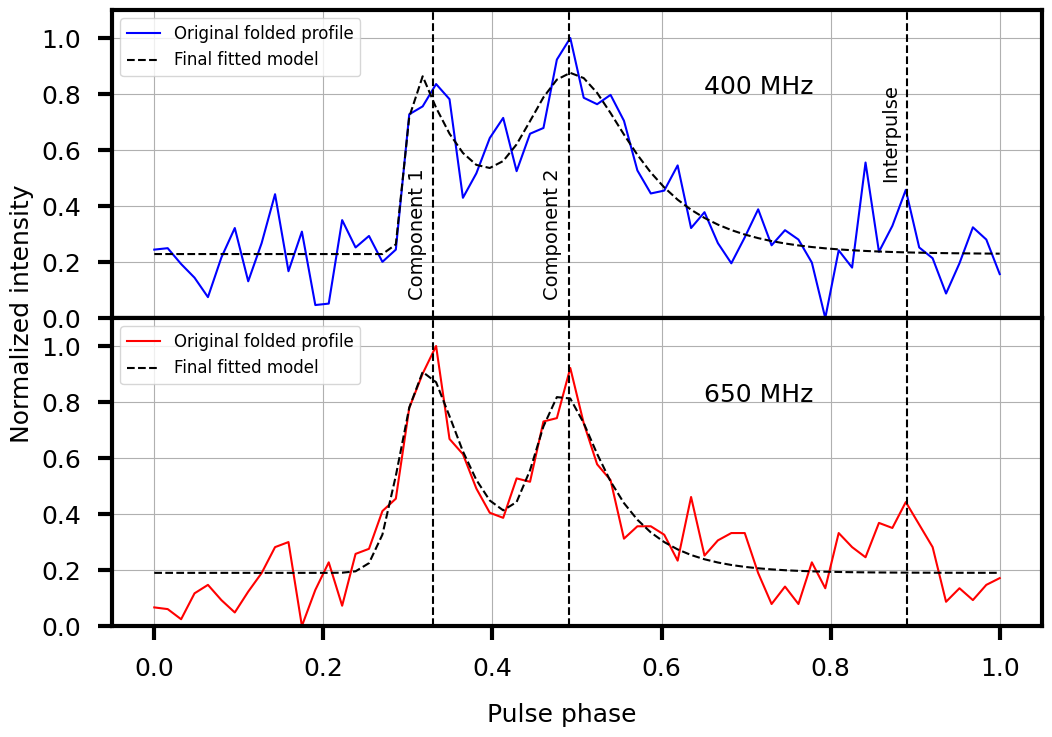}
    \caption{The coherently dedispersed folded profile of the of J1617$-$2258A for uGMRT Band-3 ($f_c = 650\:\rm MHz$) and Band-4 ($f_c = 650\:\rm MHz$). For Band-4, the sampling was 10.24 $\mu \rm s$ and for Band-3, the sampling was 40.96 $\mu \rm s$ and both foldings in the plot are downsampled to 64 bins for better visualization. The black line represents the best-fit profile according to Equation \ref{eqn:pulsar_profile_expression}.}
    \label{fig:foldedCDprofiles}
\end{figure}
The low-frequency high-time resolution pulse profile, which is unaffected by intra-channel dispersion was obtained by observing the pulsar in uGMRT Band-3 coherently-dispersed (CD) mode at the MSP DM at 40.96 $\mu$s sampling with 1024 channels. The Band-4 CD profile was obtained from the timing observation, where, along with the PA mode (used for timing), we simultaneously observed the MSP in CD mode with 10.24 $\mu$s sampling and 512 channels. This allowed us to obtain two (Band-3 and Band-4) CD pulse profiles, with reasonable SNRs for further study. We utilised these CD pulse profiles to characterise the low-frequency profile evolution and determine the scattering timescale.

As we can see in Figure \ref{fig:NewMSPplot} and in Figure \ref{fig:foldedCDprofiles}, the MSP has two clear profile components in both frequencies. It also has a faint interpulse. The pulse profile is broader at 400 $\rm MHz$ compared to 650 $\rm MHz$, where the two components are more distinctly defined. We followed the approach of \cite{McKinnon_2014} to determine the scattering timescale and intrinsic profile evolution. In this method, we fit the coherently dedispersed pulse profiles, assuming that all components intrinsically follow a Gaussian distribution and share the same scattering timescale at a given frequency. Additionally, since coherent dedispersion was applied, DM smearing is considered negligible.
According to \cite{McKinnon_2014}, for  each profile component, the scattering broadened model can be defined as:
\begin{eqnarray} \label{eqn:pulsar_profile_expression}
     f(t,\tau) & = & A + \frac{S}{2\tau} \exp\left(\frac{\sigma^2}{2\tau^2}\right)\times \exp\left(-\frac{(t-\mu)}{\tau}\right)\times \nonumber \\
        & &\left(1+ erf \left[\frac{t-(\mu+\sigma^2/\tau)}{\sigma\sqrt{2}}\right]\right)
\end{eqnarray}

Here, $A$ is the constant offset, $S$ is the integrated flux density of the component, $\mu$ is the mean, $\sigma$ is the standard deviation, $erf$ is the error function, and $\tau$ is the scattering timescale.

Both Bands' (Band-3 and Band-4) CD pulse profiles' components (except the interpulse) were fitted and components' intrinsic widths and scattering timescale were determined. For each Band, during the fitting of the multiple components, the scattering timescale ($\tau$) was kept fixed, while the components' intrinsic widths (here it's the components' HPBW: $2.355\times \sigma$) are independent of each other. The fitting was done using the Python package {\tt curve\_fit}\footnote{The package {\tt curve\_fit} can be found here: \url{https://docs.scipy.org/doc/scipy/reference/generated/scipy.optimize.curve_fit.html}}.

\begin{table}
\vspace{4mm}
\centering
\caption{The scattering timescale along with the width of both the components obtained from fitting the CD data (data unaffected by intrachannel de-dispersion). We use the convention used in \cite{McKinnon_2014} to fit our folded pulse profile. For Band-4, the profile time resolution was 10.24 $\mu \rm s$, and for Band-3, it was 40.96 $\mu \rm s$. The error presented here is obtained from the covariance matrix of the {\tt curve\_fit} algorithm.}
\begin{tabular}{|c|c|c|c|}
\toprule
\midrule
Frequency & Scattering & Component & Intrinsic \\
(MHz) & width (ms) & number & HPBW (ms) \\
\midrule
\multirow{2}{*}{400} & \multirow{2}{*}{0.328$\pm$0.104} & 1 & 0.075$\pm$0.054 \\
\cline{3-4}
    & & 2 & 0.416$\pm$0.142 \\
\midrule
\multirow{2}{*}{650} & \multirow{2}{*}{0.230$\pm$0.046} & 1 & 0.229$\pm$0.037 \\
\cline{3-4}
    & & 2 & 0.264$\pm$0.063 \\
\bottomrule
\end{tabular}
\label{table:profile}
\end{table}
In Table \ref{table:profile}, we summarise the obtained pulse widths and scattering time scales for this MSP in both the uGMRT Bands. The error we present here is the square root of the diagonal elements of the covariance matrix obtained from the {\tt curve\_fit} algorithm to get the independent fitting error of each parameter. As we see in this table (Table \ref{table:profile}), for Band-3 ($f_\mathrm{c}$ = 300 $\rm MHz$), the obtained values are not as constrained as Band-4 ($f_\mathrm{c}$ = 650 $\rm MHz$), due to very low SNR detection of the MSP (as can be seen in Figure \ref{fig:foldedCDprofiles}).

Figure \ref{fig:foldedCDprofiles} shows the averaged pulse profile for both frequency Bands, each defined by its central frequency ($f_\mathrm{c}$). The profile presented here is obtained after binning the data into 64 intervals for enhanced clarity, with the original fitted model of un-averaged data represented by the dashed curve.

\subsection{Timing} \label{subsec:timing}

Here, we present a detailed discussion of the pulsar's timing, leading to a phase-coherent timing solution, followed by an analysis of the results obtained from the timing study.

\subsubsection{Deriving the orbit} \label{subsubsection:timingmethod}

Given the detection with a significant acceleration in the discovery observation, it was clear that it is a binary MSP. To probe the nature of the binary orbit, we followed up on this MSP for seven epochs (including the discovery epoch and one confirmation epoch), with a minimum separation of one day and a maximum separation of one month in two consecutive epochs.

\begin{figure}
\centering
    \includegraphics[width=\columnwidth]{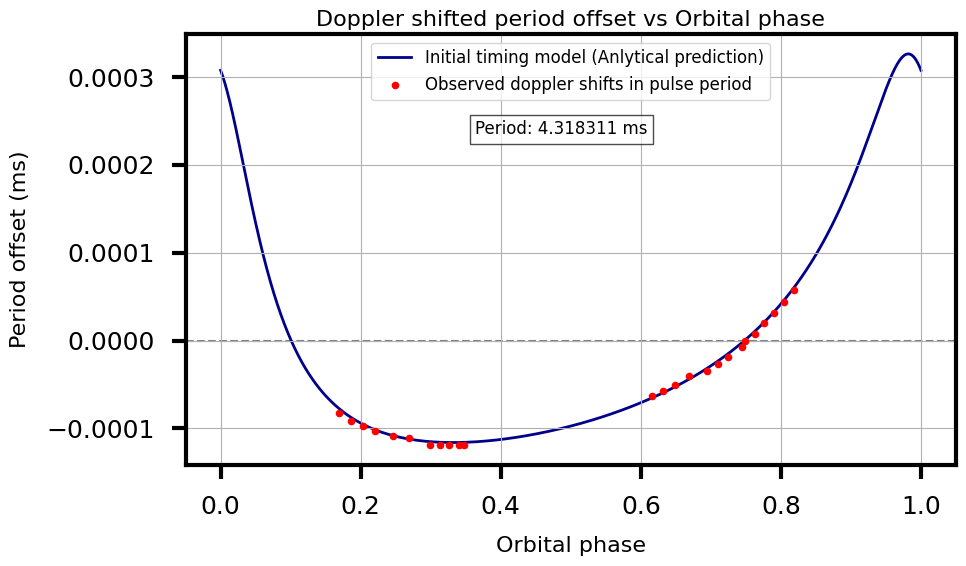}
    \caption{Doppler shifted period versus orbital phase. The scattered red points represent the period offsets we got from the detections from the observations (seven epochs) we used for the initial timing, after subtracting the Doppler shift from the Earth's motion (i.e., barycentering). The blue curve represents the best orbital model fit of the observed data produced by {\tt FITORBIT}. We can clearly see the eccentric nature of the orbit, deviating from the sinusoidal prediction for a circular orbit.}
    \label{fig:initialtimingmodelplot}
\end{figure}

\begin{figure}
\centering
    \includegraphics[width=\columnwidth]{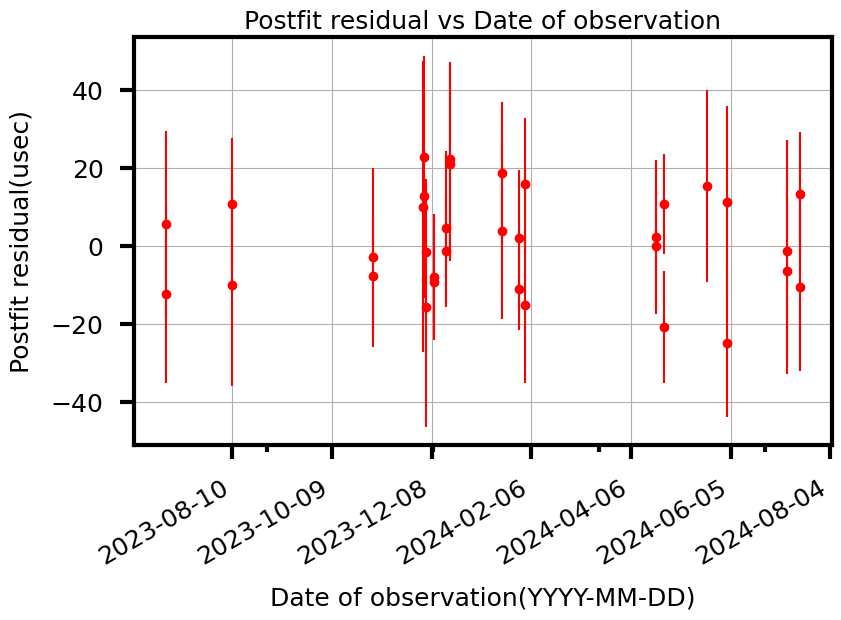}\\
    \vspace{5mm}
    \includegraphics[width=\columnwidth]{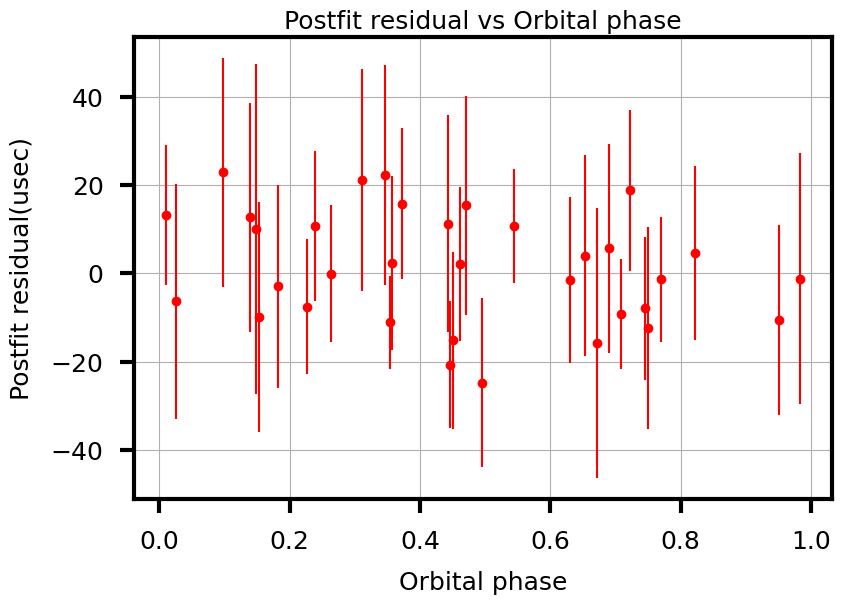}
    \caption{Post-fit timing residual versus epoch (top) and orbital phase (bottom, {\bf where the mean anomaly is measured from periastron}) after achieving the phase-coherent timing solution. The vertical lines with each point represent the rms error of each TOA.}
    \label{fig:timingresidualplots}
\end{figure}

For each epoch, we derived the pulsar period evolution with time and used the timing package {\tt FITORBIT}\footnote{The GitHub page for the {\tt FITORBIT} package: \url{https://github.com/vivekvenkris/fitorbit}} to fit a Keplerian model to the temporal variation of spin period. In Figure \ref{fig:initialtimingmodelplot} we can see the observed spin periods as a function of orbital phase and the best fit orbital model derived by {\tt FITORBIT}, where we found that J1617$-$2258A is in a highly eccentric (e $\sim$ 0.5374) relatively compact relativistic orbit having an orbital period ($P_\mathrm{b})\, \sim\, 18.94$ hours, and a projected semi-major axis $x \, = \, 0.4642\,$s. The mass function of a binary system is given by:
\begin{equation} \label{eqn:mass_function}
    f(M_\mathrm{p},M_\mathrm{c}) = \frac{(M_\mathrm{p} \sin i)^3}{(M_\mathrm{p} + M_\mathrm{c})^2} = \frac{4\pi^2}{T_{\odot}}\frac{x^3}{P_\mathrm{b}^2} = 1.7247(2) \times 10^{-4}\, \rm M_{\odot},
\end{equation}
where $T_{\odot} = ({\cal G M})^{\rm N}_{\odot}/c^3 = 4.925490947 \, \mu \rm s$ is an exact quantity, the
solar mass parameter $(\cal{G M})^{\rm N}_{\odot}$ \citep{Prsa_2016} in time units.
Assuming $M_\mathrm{p}\, = \, 1.4 \, \rm M_{\odot}$, we obtain a minimum ($i\, = \,90^\circ$) and median ($i\, = \,60^\circ$) companion masses of 0.072 M$_\odot$ and 0.084 M$_\odot$ respectively.

Although at this stage the timing model was not well constrained, the large eccentricity and relatively compact orbit made this a promising system for the precise measurement of multiple relativistic effects in the orbital motion, which will enable mass measurements in the near future.

\subsubsection{Phase connection and timing solution}
To achieve phase connection, we conducted biweekly follow-up observations of J1617$-$2258A using uGMRT Band-4 (550-750 $\rm MHz$) on 18 occasions, including discovery, confirmation, and follow-up epochs, from July 2023 to August 2024.
The timing observations are still ongoing on a monthly cadence. The average detection significance per epoch ranges from $15\sigma$ to $30\sigma$ depending on the observation duration and phasing efficiency.

We implemented a three-step approach to enhance the precision of the pulse times of arrival (ToAs) measurements and refine the timing solution. We began by using {\tt TEMPO}\footnote{TEMPO GitHub page: \url{https://github.com/nanograv/tempo}} to connect the ToAs of nearby epochs, continuing this process until no further unambiguous connections could be established. At this stage, we derived an improved ephemeris, which was then used to re-fold all data, refine pulse profile templates, and re-compute the ToAs. When no more connections could be made, we used {\tt DRACULA} \citep{Freire_Ridolfi_2018}. It is a {\tt TEMPO}-based software that iteratively determines the unique solution. Finally, with the new parameter file, we again folded all the epochs, generated relatively fewer ToAs to reduce their uncertainties, and did the final iteration of timing to obtain our unique precise timing model, which is presented
in Table~\ref{table:timing}. This uses the DE440 Solar System Ephemeris \citep{Park_2021} to account for the Earth's motion around the Solar System Barycenter and the ``DD'' orbital model \citep{DD86} to describe the pulsar's orbit. Figure \ref{fig:timingresidualplots} represents the post-fit residuals, which are the ToA values minus the prediction of the timing solution in Table~\ref{table:timing} for the same rotation.

\begin{table}[h!]
\centering
\caption{Timing solution of J1617$-$2258A}
\begin{scriptsize}
\begin{tabular}{ll}
\toprule
\midrule
Parameters of observation and timing solution \\
\midrule
PSR \dotfill & J1617$-$2258A \\
Reference epoch (MJD) \dotfill & 60166.864788 \\
Start of timing data (MJD) \dotfill & 60126.783 \\
End of timing data (MJD) \dotfill & 60508.571 \\
Number of ToAs \dotfill & 34 \\
Postfit timing residuals, RMS ($\mu$s) \dotfill & 12.137 \\
Solar system ephemeris \dotfill & DE440 \\
Time units \dotfill & TDB \\
Binary model \dotfill & DD \\
\toprule
Measured parameters \\
\midrule
Right Ascension, $\alpha$ (J2000) \dotfill & 16:17:02.821(4) \\
Declination, $\delta$ (J2000) \dotfill & $-$22:58:36.8(3) \\
Spin frequency, $f$ (Hz) \dotfill & 231.572028561(2) \\
First spin frequency derivative, $\dot{f}$ (Hz s$^{-1}$) \dotfill & $-$3.3(1)$\times$10$^{-15}$ \\
Dispersion measure, DM (pc cm$^{-3}$) \dotfill & 66.812 \\
Orbital period, $P_\mathrm{b}$ (d) \dotfill & 0.78914809(7) \\
Projected semi-major axis, $x$ (lt-s) \dotfill & 0.46421(1) \\
Eccentricity, $e$ \dotfill & 0.53744(3) \\
Time of periastron, $T_0$ (MJD) \dotfill & 60166.84499(2) \\
Longitude of periastron, $\omega$ (degrees) \dotfill & 24.703(9) \\
Rate of periastron advance, $\dot{\omega}$ (deg year$^{-1}$) \dotfill & 0.5851 $\pm$ 0.0149 \\
\toprule
Derived parameters \\
\midrule
Position offset from GC centre ($''$) \dotfill &  6.444 \\
Position offset from GC centre (core radii) \dotfill &  0.72 \\
Spin period, $P$ (ms) \dotfill & 4.318310834921(3) \\
Spin period derivative, $\dot{P}$ (s s$^{-1}$) \dotfill & 6.1(2)$\times$10$^{-20}$ \\
Characteristic age, $\tau_c$ (Gyr) \dotfill & $> 0.25$ \\
Estimated surface magnetic field, B ($10^9$ G) \dotfill & $< 1.1$ \\
Mass function, $f(M_\mathrm{p},M_\mathrm{c})$ (M$_\odot$) \dotfill &  0.00017247(2)\\
The total mass of the system, $M_\mathrm{tot}$ (M$_\odot$) \dotfill & 1.67$\pm$0.06 \\
Minimum companion mass, $M_\mathrm{c, min}$ (M$_\odot$) \dotfill & 0.07203(3) \\
Median companion mass, $M_\mathrm{c, min}$ (M$_\odot$) \dotfill & 0.08361(3) \\
Maximum pulsar mass, $M_\mathrm{p, max}$ (M$_\odot$) \dotfill & 1.60$\pm$0.06 \\
\bottomrule
\end{tabular}
\end{scriptsize}
\label{table:timing}
\end{table}

\subsubsection{Position and spin parameters} 

The precise position of J1617$-$2258A obtained from the timing (listed in Table~\ref{table:timing}) coincides with the radio source (as seen in Figure \ref{fig:NewMSPimage}) discussed in Section \ref{subsec:localization}, confirming its association with J1617$-$2258A.
The pulsar is located 6.44$''$ from the cluster centre defined by \cite{Vasiliev_2021}. This distance represents about 0.72 core radii, where the core radius is assumed to be $\sim$ 9$''$, see \citet[][2010 edition]{Harris_1996}. At the distance of M80, about 9.8 kpc \cite{Vasiliev_2021}, this represents a projected distance of 0.306 pc.

The observed spin period derivative of the pulsar ($\dot{P}_\mathrm{obs}$) is the sum of the intrinsic spin period derivative ($\dot{P}_\mathrm{int}$) plus contributions from the acceleration of the system in the Galaxy ($a_\mathrm{Gal}$, from which we subtract the acceleration of the Solar System in the Galaxy) and in the GC ($a_\mathrm{GC}$), both projected along the line of sight to the system and finally the Shklovskii effect \citep{Shklovskii_1970}, which can be derived from the total proper motion of the system ($\mu$)and its distance ($d$):
\begin{equation} \label{eqn:Pdot_expression}
\frac{\dot{P}_\mathrm{obs}}{P} = \frac{\dot{P}_\mathrm{int}}{P} + \frac{a_\mathrm{Gal}}{c} + \frac{a_\mathrm{GC}}{c} + \frac{\mu^2 d}{c}
\end{equation}
To calculate $a_\mathrm{Gal}$, we have used the Galactic potential model of \cite{McMillan2017}, which yields $a_\mathrm{Gal} = -0.409 \times 10^{-9} \rm \, m\, s^{-2}$. To calculate the Shklovskii effect, we have assumed that the proper motion of the pulsar (J1617$-$2258A) is the same as that of the GC (NGC~6093), which is 
$\mu_\alpha = -2.934\, \pm \, 0.027 \, \rm mas \, yr^{-1}$ and $\mu_\delta = -5.578\, \pm \, 0.026 \, \rm mas \, yr^{-1}$ \citep{Vasiliev_2021}; this results in an apparent acceleration of $a_\mathrm{Shk} = \mu^2 d = +0.284 \times 10^{-9} \rm \, m\, s^{-2}$. Thus the sum of the kinematic contribution from the Galaxy is 
$-0.122 \times 10^{-9} \rm \, m\, s^{-2}$.
From this, we can derive, for $\dot{P}_\mathrm{int} = 0$, an upper limit of the acceleration of the pulsar in the gravitational field of M80 (projected along the line of sight) as:
\begin{equation} \label{eqn:GC_acceleartion_expression}
a_\mathrm{l, max} = \frac{\dot{P}_\mathrm{obs}}{P}c - a_\mathrm{Gal} - a_\mathrm{Shk} = +4.35\, \times \, 10^{-9} \, \rm m\, s^{-2}.
\end{equation}
To model the gravitational field of M80, we use the analytical model described by \cite{Freire_2005}, where we use a velocity dispersion in the core of 12.4 km s$^{-1}$ \citep[][2010 edition]{Harris_1996}.
For the line of sight of the pulsar, this model predicts a maximum/minimum cluster acceleration of $a_\mathrm{GC} = \pm 14.75 \times 10^{-9} \rm \, m\, s^{-2}$, which is significantly larger than
$a_\mathrm{l, max}$ of the pulsar; this means that the value of $a_\mathrm{l, max}$ is well within the range of accelerations predicted by our mass model for M80. Assuming the minimum possible cluster acceleration, we obtain an absolute upper limit of $\dot{P}_\mathrm{int} < 2.75 \times 10^{-19}$. From this value, the limits on characteristic age ($\tau$), and surface magnetic filed strength ($B$) can be estimated using:
\begin{equation} \label{eqn:charact_age}
    \tau_c = P/2\dot{P},\:B = 3.2\times10^{19}\left(P\dot{P}\right)^{1/2},
\end{equation}
from which we obtain, $\tau_c > 0.25\, \rm Gyr$, and $B\, < \, 1.1\, \times\, 10^9\, \rm G$. These values are not very constraining, but they are consistent with typical values obtained for MSPs in the Galactic disk. Much improved values will be obtained when the orbital period derivative of the system becomes measurable (see e.g., discussion in \cite{Freire_2017}).

\subsection{Binary characteristics}

\label{subsec:characteristics}

In this section, we discuss the binary characteristics of J1617$-$2258A, the unique compact eccentric orbit and its companion type.

\subsubsection{The compact eccentric orbit of J1617$-$2258A} \label{subsubsec:compactorbit}
The evolution of the LMXBs greatly shapes the resultant MSP system and its properties:
the transfer of matter from the extended companion to the NS spins it up, decreases its magnetic field and circularises the orbit via tidal interactions \citep{Tauris_2023}. After the end of this mass transfer, the NS becomes an MSP in different systems like NS-WD binaries, or eclipsing systems like redback or black-widow MSPs (see \citealt{Jia_2016}), all of which have nearly circular orbits.

\begin{figure*}
\centering
    \includegraphics[width=0.7\textwidth]{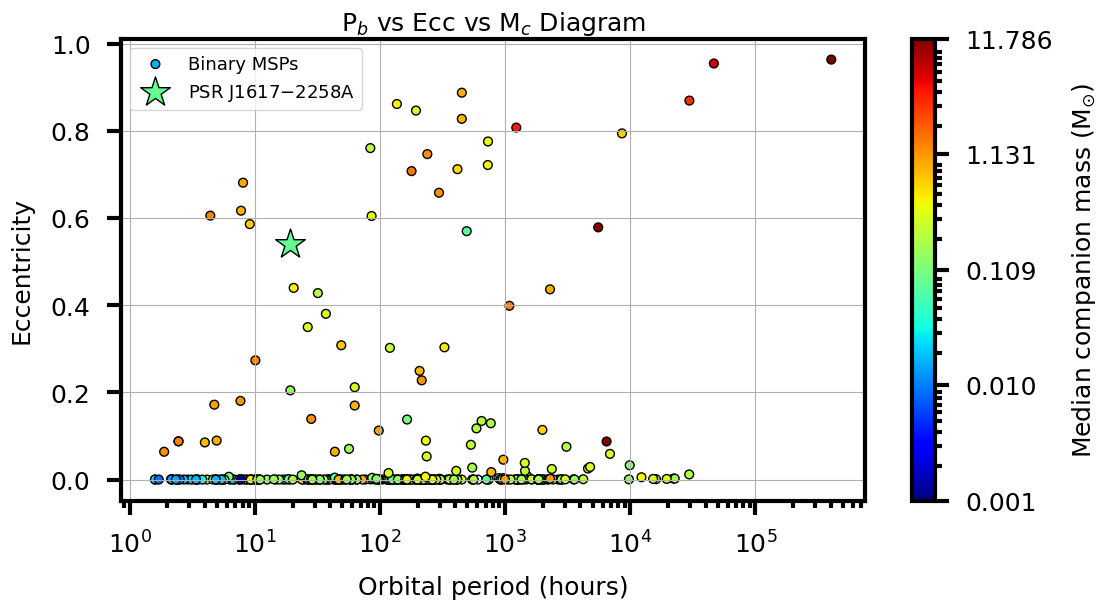}
    \caption{The plot of the orbital period versus eccentricity for all the binary MSPs detected till now, along with GCGPS's J1617$-$2258A. All the binary MSP data were taken from the ATNF Pulsar catalogue \citep{Manchester_2025}. The additional axis is the median companion mass of each system represented as the colour of each dot, and the colour bar represents the range. As seen in the plot, highly eccentric systems generally have either wide orbits, high-mass companions, or a combination of both. In contrast, the newly discovered GC MSP J1617$-$2258A, having a compact orbit and a companion with low median mass, is quite eccentric and sits in a unique spot in the plot. Only four more systems are more eccentric while having a more compact orbit, but interestingly, all of them have median companion masses higher than J1617$-$2258A.}
    \label{fig:PbvsMcvseccplot}
\end{figure*}

Interestingly in our case, J1617$-$2258A does not follow this trend: it is in a $\sim$ 18.94 hours compact but highly eccentric (e $\sim$ 0.54) binary orbit, which is quite uncommon. In Figure \ref{fig:PbvsMcvseccplot}, the unique position of this MSP is visible, where there are only 4 MSPs (3 are from the GCs) orbits more compact and more eccentric than this orbit, but all of them have a median companion mass higher than J1617$-$2258A's median companion mass.

The reason for J1617$-$2258A's highly eccentric compact orbit is likely to be due to past close flybys with other GC stars, which can change the eccentricity of the orbit \citep[e.g.][]{Phinney_1992}. An exchange encounter is also possible, but such encounters generally replace the light companions with more massive objects \citep{Freire2004, Lynch2012, DeCesar_2015, Barr_2024}.

\subsubsection{Rate of advance of periastron}
\label{total_mass_estimation}
With more than one year of timing baseline, this highly eccentric compact orbit allowed us to detect one post-Keplerian (PK) parameter, the rate of advance of periastron ($\dot{\omega}$) in our timing with sufficient detection significance.
If this is purely relativistic, then in general relativity (GR), this is related to the total mass of the system by:
\begin{equation} \label{eqn:wdotexpression}
    \dot{\omega} = 3T_{\odot}^{2/3}\left(\frac{P_\mathrm{b}}{2\pi}\right)^{-5/3}\frac{M_{\rm tot}^{2/3}}{1-e^2},
\end{equation}
where $M_\mathrm{tot} = M_\mathrm{p} + M_\mathrm{c}$.
Our measurement of $\dot{\omega}$ results in $M_\mathrm{tot} = 1.67\, \pm \, 0.06 \, \rm M_{\odot}$. In Fig.~\ref{fig:mass_mass}, this constraint and its 1-$\sigma$ uncertainty are indicated by the solid red lines.

If we combine the nominal value of the total mass with the mass function equation (given by eq.~\ref{eqn:GC_acceleartion_expression}, which excludes the grey area in the right plot of Fig.~\ref{fig:mass_mass}), we obtain for $i = 90^\circ$ a minimum companion mass of 0.072~M$_{\odot}$ and a maximum pulsar mass of 1.60~M$_{\odot}$. A detailed statistical Bayesian approach for the individual mass measurement is done on Section \ref{subsec:massestimates}.

\subsubsection{Probability distribution of pulsar mass}
\label{subsec:massestimates}

\begin{figure*}
\centering
    \includegraphics[width=0.8\textwidth]{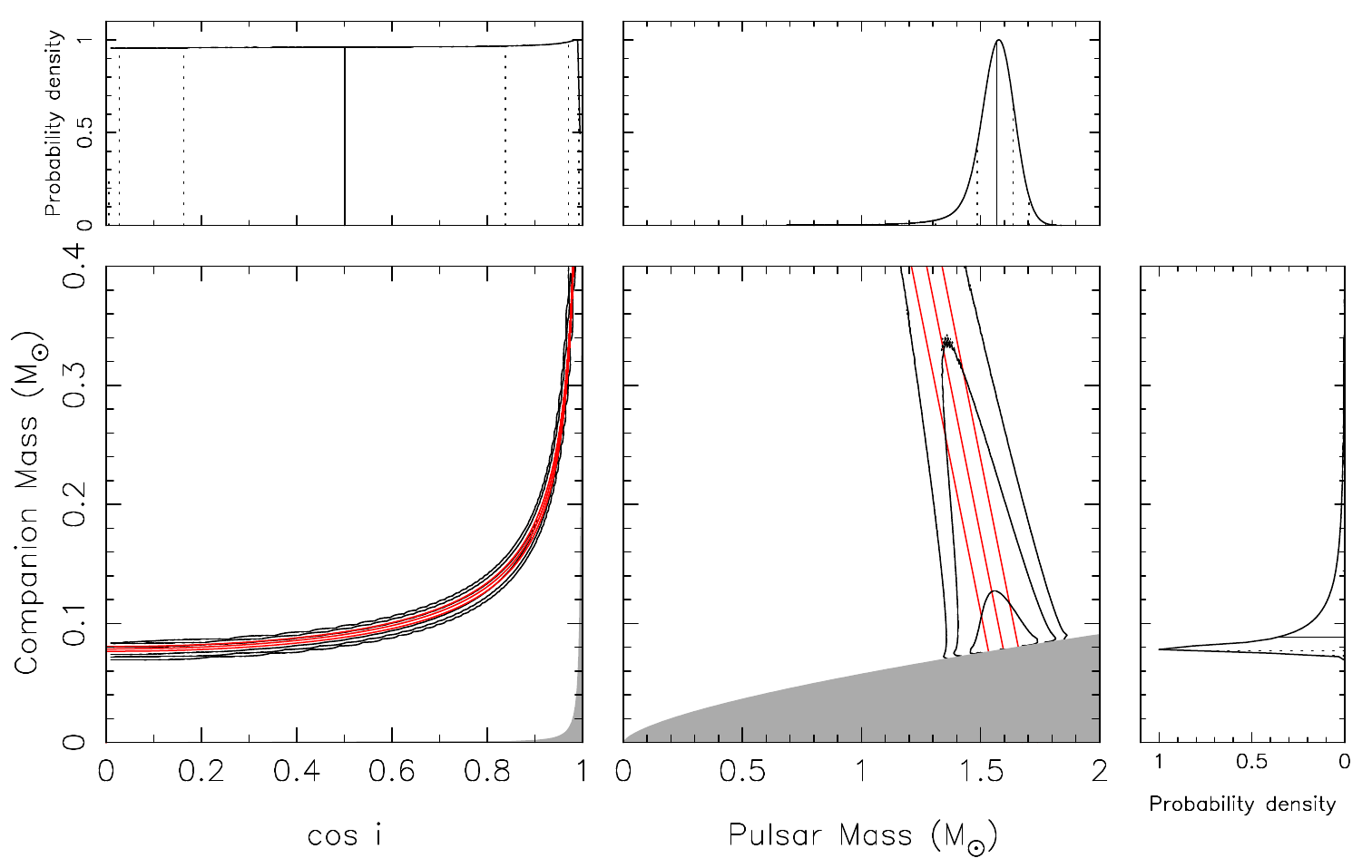}
    \caption{Mass-mass diagram for J1617$-$2258A. {\em Main left panel}: $\cos i$ - $M_\mathrm{c}$ space. The grey area is excluded by the requirement that $M_\mathrm{p} > 0$. {\em Main right panel}: $M_\mathrm{p}$ - $M_\mathrm{c}$ space. The grey area is excluded by the requirement that $\sin i \leq 0$. The mass constraints from the measurement of $\dot{\omega}$ are depicted in red. The solid black contours include 68.3\%, 95.4\% and 99.7\% of all probability in the pdfs in each panel. {\em Side panels, top left:} 1-D pdf for $\cos i$, {\em Top right: }1-D pdf for $M_\mathrm{p}$, {\em right-side panel:} 1-D pdf for $M_\mathrm{c}$, the solid and dotted lines depict their medians and 1, 2 and 3-$\sigma$ equivalent percentiles.}
    \label{fig:mass_mass}
\end{figure*}

Given the only measurement of $\dot{\omega}$, and the unavailability of significant detection of other relativistic parameters, in the previous section, we were only able to derive limits on the pulsar and the companion mass based on the total mass of the system. To better quantify the masses and their uncertainties given the constraints of the timing, and to properly include the constraints from the lack of measurement of the Shapiro delay and any other relativistic effects that might be present in the timing with low significance, we have made a Bayesian $\chi^2$ map. This samples $\cos i$ and $M_\mathrm{tot}$ uniformly. For each point in this grid, we calculate $M_\mathrm{c}$ and then include the values of $M_\mathrm{tot}$ and $M_\mathrm{c}$ in a DDGR \citep{Taylor_1989} timing solution, where they are kept fixed. This type of orbital model assumes the validity of GR in a self-consistent way takes into account all relativistic constraints that might be present in the timing. We then use {\tt TEMPO} to obtain the optimal solution for those values, recording the value of $\chi^2$. This is then converted into a 2-D probability density function (pdf) using the Bayesian method of \cite{Splaver_2002}. We then project this pdf into the $\cos i$ - $M_\mathrm{c}$ and $M_\mathrm{p}$ - $M_\mathrm{c}$ spaces, and from these we derive the 1-D pdfs for $M_\mathrm{tot}$, $M_\mathrm{p}$, $M_\mathrm{c}$ and $\cos i$.

These 2-D and 1-D pdfs are depicted in Fig.~\ref{fig:mass_mass}.
In the main panels, the pdfs are depicted by contours that include 68.3\%, 95.4\% and 99.7\% of the 2-D pdf in each panel. In the marginal panels, we depict the 1-D pdfs for $\cos i$, $M_\mathrm{p}$ and $M_\mathrm{c}$.

The medians and $\pm 1$-$\sigma$ percentiles of these pdfs are:
$M_\mathrm{tot} \, = \, 1.672 \pm 0.063 \, \rm M_{\odot}$,
$M_\mathrm{p} \, = \, 1.567^{+0.060}_{-0.080} \, \rm M_{\odot}$ and
$M_\mathrm{c} \, = \, 0.086^{+0.056}_{-0.012} \, \rm M_{\odot}$.

For $\cos i$, we see that the pdf is nearly flat. This means that no other relativistic effects are detected and that at the moment, the individual masses cannot be estimated, other than via the statistical assumption of a flat distribution of $\cos i$.

\subsubsection{The companion and binary type}

\label{subsubsec:companion}

During timing observation, we roughly sampled every part of the orbit (Figure \ref{fig:timingresidualplots}) and for every observation, this MSP was detected at about the expected SNR and no eclipsing phenomenon was observed. In most cases, spider MSPs (redbacks and black widows) show eclipsing due to the interaction of the pulsar pulse with its non-degenerate companion outflows \citep{Yan_2021, Kumari_2024}. If this MSP system is a spider system then either the eclipsing is covering a very small part of the orbital phase and the region remained unsampled till now or the eclipsing could be frequency dependent, but for that fine sampling of the orbital phase and observation in different frequency except 550$-$750 $\rm MHz$ is needed to confirm if it is a spider MSP system.

Another possibility is that the system is not a spider system but rather a neutron star-white-dwarf (NS$-$WD) binary system. For MSP with a Population-I WD companion, according to \cite{Tauris_1999}'s $P_\mathrm{b}-M_\mathrm{c}$ correlation, for compact orbit($P_\mathrm{b}\:\sim\:1\:\rm days$) we can write:
\begin{equation} \label{eqn:Pb_Mc_correlation}
    \frac{M_\mathrm{c}}{\, \rm M_{\odot}} = \left(\frac{P_\mathrm{b}}{b}\right)^{1/a} + c
\end{equation}
Where, $P_\mathrm{b}$ in days, a = 4.50, b = 1.2 $\times\:10^5$, and c= 0.12. Putting our $P_\mathrm{b}$ value into Equation \ref{eqn:Pb_Mc_correlation}, we get our $M_\mathrm{c}\, = \, 0.19 \, \rm M_{\odot}$, which would imply $M_\mathrm{p}\, = \, 1.48 \pm 0.6 \, \rm M_{\odot}$ and $i \sim 24^\circ$. If this low orbital inclination is correct, measuring the Shapiro delay in this system will not be possible, which agrees with our previous analysis.

The relatively low companion mass of J1617$-$2258A is compatible with its current companion being the white dwarf remnant of the star that recycled the pulsar. Further study is needed to determine its companion type.

\section{SUMMARY} \label{sec:summary}

This paper discusses the Globular Clusters GMRT Pulsar Search (GCGPS) survey. GCGPS has recently started and is currently searching for MSPs in GCs. This survey is designed to search for GC pulsars in the low-frequency range (300$-$750 MHz). The target selection criteria and observation strategy are optimized to increase discovery potential. GCGPS survey exploits uGMRT's Y-shaped array to form two beams with different baseline lengths to optimize sky coverage along with beam sensitivity.

We present the discovery of the first MSP in the GC NGC6093 (M80). We localise the MSP from imaging of the field as well as from timing solutions at sub-arc-sec precession. The timing of the MSP revealed that it is in a highly eccentric ($e\:\sim\:0.54$) compact ($P_\mathrm{b}\:\sim\:18.94\,$h) relativistic binary. Measurement of one PK parameter (advance of periastron, $\dot{\omega}$) leads to measurement of the total mass of the system ($M_\mathrm{tot} = 1.67\pm0.06\:\, \rm M_{\odot}$), this and the mass function imply $M_\mathrm{c} \geq 0.072~ \rm M_{\odot}$ and $M_\mathrm{p} \leq 1.60
\rm M_{\odot}$. The limit on the companion mass is consistent with it being a He WD whose progenitor spun up the pulsar, although we cannot exclude the possibility that the system might have formed in an exchange interaction. There are only four such systems that are more compact and eccentric than this system, but all have a higher median companion mass than this system.

\begin{acknowledgments}
We acknowledge the support of the Department of Atomic Energy, Government of India, under project No. 12-R\&D-TFR5.02-0700. The GMRT is run by the National Center for Radio Astrophysics of the Tata Institute of Fundamental Research, India. We thank the uGMRT operators for their coordinated effort in conducting the GCGPS survey observations. PCCF gratefully acknowledges continuing support from the Max-Planck-Gesellschaft and hospitality of the Academia Sinica Institute of Astronomy and Astrophysics in Taipei, Taiwan, where part of this work was conducted while he was a Visiting Scholar. We also acknowledge support from the \textit{Building Indo-UK Collaborations towards the SKA} program, which facilitated the development of the pulsar search pipeline, particularly the highly efficient GPU-based version. We also thank our anonymous referee for his/her insightful comments, which helped make the necessary corrections.
\end{acknowledgments}

\bibliography{bibliography}{}

\begin{thebibliography}{}
\expandafter\ifx\csname natexlab\endcsname\relax\def\natexlab#1{#1}\fi
\providecommand{\url}[1]{\href{#1}{#1}}
\providecommand{\dodoi}[1]{doi:~\href{http://doi.org/#1}{\nolinkurl{#1}}}
\providecommand{\doeprint}[1]{\href{http://ascl.net/#1}{\nolinkurl{http://ascl.net/#1}}}
\providecommand{\doarXiv}[1]{\href{https://arxiv.org/abs/#1}{\nolinkurl{https://arxiv.org/abs/#1}}}

\bibitem[{{Abbate} {et~al.}(2022){Abbate}, {Ridolfi}, {Barr}, {Buchner}, {Burgay}, {Champion}, {Chen}, {Freire}, {Gautam}, {Grie{\ss}meier}, {K{\"u}nkel}, {Kramer}, {Padmanabh}, {Possenti}, {Ransom}, {Serylak}, {Stappers}, {Venkatraman Krishnan}, {Behrend}, {Breton}, {Levin}, \& {Men}}]{Abbate2022}
{Abbate}, F., {Ridolfi}, A., {Barr}, E.~D., {et~al.} 2022, \mnras, 513, 2292, \dodoi{10.1093/mnras/stac1041}

\bibitem[{{Abbate} {et~al.}(2023){Abbate}, {Ridolfi}, {Freire}, {Padmanabh}, {Balakrishnan}, {Buchner}, {Zhang}, {Kramer}, {Stappers}, {Barr}, {Chen}, {Champion}, {Ransom}, \& {Possenti}}]{Abbate_2023}
{Abbate}, F., {Ridolfi}, A., {Freire}, P.~C.~C., {et~al.} 2023, \aap, 680, A47, \dodoi{10.1051/0004-6361/202347725}

\bibitem[{{Abdo} {et~al.}(2010){Abdo}, {Ackermann}, {Ajello}, {Baldini}, {Ballet}, {Barbiellini}, {Bastieri}, {Bellazzini}, {Blandford}, {Bloom}, {Bonamente}, {Borgland}, {Bouvier}, {Brandt}, {Bregeon}, {Brigida}, {Bruel}, {Buehler}, {Buson}, {Caliandro}, {Cameron}, {Caraveo}, {Carrigan}, {Casandjian}, {Charles}, {Chaty}, {Chekhtman}, {Cheung}, {Chiang}, {Ciprini}, {Claus}, {Cohen-Tanugi}, {Conrad}, {Decesar}, {Dermer}, {de Palma}, {Digel}, {Silva}, {Drell}, {Dubois}, {Dumora}, {Favuzzi}, {Fortin}, {Frailis}, {Fukazawa}, {Fusco}, {Gargano}, {Gasparrini}, {Gehrels}, {Germani}, {Giglietto}, {Giordano}, {Glanzman}, {Godfrey}, {Grenier}, {Grondin}, {Grove}, {Guillemot}, {Guiriec}, {Hadasch}, {Harding}, {Hays}, {Jean}, {J{\'o}hannesson}, {Johnson}, {Johnson}, {Kamae}, {Katagiri}, {Kataoka}, {Kerr}, {Kn{\"o}dlseder}, {Kuss}, {Lande}, {Latronico}, {Lee}, {Lemoine-Goumard}, {Llena Garde}, {Longo}, {Loparco}, {Lovellette}, {Lubrano}, {Makeev}, {Mazziotta}, {Michelson}, {Mitthumsiri}, {Mizuno}, {Monte}, {Monzani},
  {Morselli}, {Moskalenko}, {Murgia}, {Naumann-Godo}, {Nolan}, {Norris}, {Nuss}, {Ohsugi}, {Omodei}, {Orlando}, {Ormes}, {Pancrazi}, {Parent}, {Pepe}, {Pesce-Rollins}, {Piron}, {Porter}, {Rain{\`o}}, {Rando}, {Reimer}, {Reimer}, {Reposeur}, {Ripken}, {Romani}, {Roth}, {Sadrozinski}, {Saz Parkinson}, {Sgr{\`o}}, {Siskind}, {Smith}, {Spinelli}, {Strickman}, {Suson}, {Takahashi}, {Takahashi}, {Tanaka}, {Thayer}, {Thayer}, {Tibaldo}, {Torres}, {Tosti}, {Tramacere}, {Uchiyama}, {Usher}, {Vasileiou}, {Venter}, {Vilchez}, {Vitale}, {Waite}, {Wang}, {Webb}, {Winer}, {Yang}, {Ylinen}, {Ziegler}, \& {Fermi LAT Collaboration}}]{Adbo_2010}
{Abdo}, A.~A., {Ackermann}, M., {Ajello}, M., {et~al.} 2010, \aap, 524, A75, \dodoi{10.1051/0004-6361/201014458}

\bibitem[{{Bahramian} {et~al.}(2013){Bahramian}, {Heinke}, {Sivakoff}, \& {Gladstone}}]{Bahramian_2013}
{Bahramian}, A., {Heinke}, C.~O., {Sivakoff}, G.~R., \& {Gladstone}, J.~C. 2013, \apj, 766, 136, \dodoi{10.1088/0004-637X/766/2/136}

\bibitem[{{Barr} {et~al.}(2024){Barr}, {Dutta}, {Freire}, {Cadelano}, {Gautam}, {Kramer}, {Pallanca}, {Ransom}, {Ridolfi}, {Stappers}, {Tauris}, {Venkatraman Krishnan}, {Wex}, {Bailes}, {Behrend}, {Buchner}, {Burgay}, {Chen}, {Champion}, {Chen}, {Corongiu}, {Geyer}, {Men}, {Padmanabh}, \& {Possenti}}]{Barr_2024}
{Barr}, E.~D., {Dutta}, A., {Freire}, P. C.~C., {et~al.} 2024, Science, 383, 275, \dodoi{10.1126/science.adg3005}

\bibitem[{{Becker} \& {Tr{\"u}mper}(1999)}]{Becker_1999}
{Becker}, W., \& {Tr{\"u}mper}, J. 1999, \aap, 341, 803, \dodoi{10.48550/arXiv.astro-ph/9806381}

\bibitem[{{Bhat} {et~al.}(2004){Bhat}, {Cordes}, {Camilo}, {Nice}, \& {Lorimer}}]{Bhat2004}
{Bhat}, N.~D.~R., {Cordes}, J.~M., {Camilo}, F., {Nice}, D.~J., \& {Lorimer}, D.~R. 2004, \apj, 605, 759, \dodoi{10.1086/382680}

\bibitem[{{Bhattacharyya} {et~al.}(2016){Bhattacharyya}, {Cooper}, {Malenta}, {Roy}, {Chengalur}, {Keith}, {Kudale}, {McLaughlin}, {Ransom}, {Ray}, \& {Stappers}}]{Bhattacharyya_2016}
{Bhattacharyya}, B., {Cooper}, S., {Malenta}, M., {et~al.} 2016, \apj, 817, 130, \dodoi{10.3847/0004-637X/817/2/130}

\bibitem[{{Bhattacharyya} {et~al.}(2019){Bhattacharyya}, {Roy}, {Stappers}, {Johnson}, {Ilie}, {Lyne}, {Malenta}, {Weltevrede}, {Chengalur}, {Cooper}, {Kaur}, {Keith}, {Kerr}, {Kudale}, {McLaughlin}, {Ransom}, \& {Ray}}]{Bhattacharyya_2019}
{Bhattacharyya}, B., {Roy}, J., {Stappers}, B.~W., {et~al.} 2019, \apj, 881, 59, \dodoi{10.3847/1538-4357/ab2bf3}

\bibitem[{{Bogdanov} {et~al.}(2011){Bogdanov}, {van den Berg}, {Servillat}, {Heinke}, {Grindlay}, {Stairs}, {Ransom}, {Freire}, {B{\'e}gin}, \& {Becker}}]{Bogdanov_2011}
{Bogdanov}, S., {van den Berg}, M., {Servillat}, M., {et~al.} 2011, \apj, 730, 81, \dodoi{10.1088/0004-637X/730/2/81}

\bibitem[{{Camilo} {et~al.}(2000){Camilo}, {Lorimer}, {Freire}, {Lyne}, \& {Manchester}}]{Camilo2000}
{Camilo}, F., {Lorimer}, D.~R., {Freire}, P., {Lyne}, A.~G., \& {Manchester}, R.~N. 2000, \apj, 535, 975, \dodoi{10.1086/308859}

\bibitem[{{Chen} {et~al.}(2023){Chen}, {Cadelano}, {Pallanca}, {Ferraro}, {Lanzoni}, {Istrate}, {Burgay}, {Freire}, {Gautam}, {Possenti}, \& {Ridolfi}}]{Chen_2023}
{Chen}, J., {Cadelano}, M., {Pallanca}, C., {et~al.} 2023, \apj, 948, 84, \dodoi{10.3847/1538-4357/acc583}

\bibitem[{{Cordes} \& {Lazio}(2002)}]{Cordes_2001}
{Cordes}, J.~M., \& {Lazio}, T.~J.~W. 2002, arXiv e-prints, astro, \dodoi{10.48550/arXiv.astro-ph/0207156}

\bibitem[{{Corongiu} {et~al.}(2024){Corongiu}, {Ridolfi}, {Abbate}, {Bailes}, {Possenti}, {Geyer}, {Manchester}, {Kramer}, {Freire}, {Burgay}, {Buchner}, \& {Camilo}}]{Corongiu_2024}
{Corongiu}, A., {Ridolfi}, A., {Abbate}, F., {et~al.} 2024, \apj, 972, 198, \dodoi{10.3847/1538-4357/ad5e74}

\bibitem[{{D'Amico} {et~al.}(2003){D'Amico}, {Possenti}, {Manchester}, {Lyne}, {Camilo}, \& {Sarkissian}}]{Amico2003}
{D'Amico}, N., {Possenti}, A., {Manchester}, R.~N., {et~al.} 2003, in Astronomical Society of the Pacific Conference Series, Vol. 302, Radio Pulsars, ed. M.~{Bailes}, D.~J. {Nice}, \& S.~E. {Thorsett}, 375

\bibitem[{{Damour} \& {Deruelle}(1986)}]{DD86}
{Damour}, T., \& {Deruelle}, N. 1986, Annales de L'Institut Henri Poincare Section (A) Physique Theorique, 44, 263

\bibitem[{{DeCesar} {et~al.}(2015){DeCesar}, {Ransom}, {Kaplan}, {Ray}, \& {Geller}}]{DeCesar_2015}
{DeCesar}, M.~E., {Ransom}, S.~M., {Kaplan}, D.~L., {Ray}, P.~S., \& {Geller}, A.~M. 2015, \apjl, 807, L23, \dodoi{10.1088/2041-8205/807/2/L23}

\bibitem[{{Dewey} {et~al.}(1985){Dewey}, {Taylor}, {Weisberg}, \& {Stokes}}]{Dewey_1985}
{Dewey}, R.~J., {Taylor}, J.~H., {Weisberg}, J.~M., \& {Stokes}, G.~H. 1985, \apjl, 294, L25, \dodoi{10.1086/184502}

\bibitem[{{Douglas} {et~al.}(2022){Douglas}, {Padmanabh}, {Ransom}, {Ridolfi}, {Freire}, {Krishnan}, {Barr}, {Pallanca}, {Cadelano}, {Possenti}, {Stairs}, {Hessels}, {DeCesar}, {Lynch}, {Bailes}, {Burgay}, {Champion}, {Karuppusamy}, {Kramer}, {Stappers}, \& {Vleeschower}}]{Douglas_2022}
{Douglas}, A., {Padmanabh}, P.~V., {Ransom}, S.~M., {et~al.} 2022, \apj, 927, 126, \dodoi{10.3847/1538-4357/ac4744}

\bibitem[{{Freire} {et~al.}(2004){Freire}, {Gupta}, {Ransom}, \& {Ishwara-Chandra}}]{Freire2004}
{Freire}, P.~C., {Gupta}, Y., {Ransom}, S.~M., \& {Ishwara-Chandra}, C.~H. 2004, \apjl, 606, L53, \dodoi{10.1086/421085}

\bibitem[{{Freire} {et~al.}(2005){Freire}, {Hessels}, {Nice}, {Ransom}, {Lorimer}, \& {Stairs}}]{Freire_2005}
{Freire}, P. C.~C., {Hessels}, J. W.~T., {Nice}, D.~J., {et~al.} 2005, \apj, 621, 959, \dodoi{10.1086/427748}

\bibitem[{{Freire} {et~al.}(2008){Freire}, {Ransom}, {B{\'e}gin}, {Stairs}, {Hessels}, {Frey}, \& {Camilo}}]{Freire2008}
{Freire}, P. C.~C., {Ransom}, S.~M., {B{\'e}gin}, S., {et~al.} 2008, \apj, 675, 670, \dodoi{10.1086/526338}

\bibitem[{{Freire} \& {Ridolfi}(2018)}]{Freire_Ridolfi_2018}
{Freire}, P. C.~C., \& {Ridolfi}, A. 2018, \mnras, 476, 4794, \dodoi{10.1093/mnras/sty524}

\bibitem[{{Freire} \& {Wex}(2024)}]{Freire_Wex_2024}
{Freire}, P. C.~C., \& {Wex}, N. 2024, Living Reviews in Relativity, 27, 5, \dodoi{10.1007/s41114-024-00051-y}

\bibitem[{{Freire} {et~al.}(2011){Freire}, {Abdo}, {Ajello}, {Allafort}, {Ballet}, {Barbiellini}, {Bastieri}, {Bechtol}, {Bellazzini}, {Blandford}, {Bloom}, {Bonamente}, {Borgland}, {Brigida}, {Bruel}, {Buehler}, {Buson}, {Caliandro}, {Cameron}, {Camilo}, {Caraveo}, {Cecchi}, {{\c{C}}elik}, {Charles}, {Chekhtman}, {Cheung}, {Chiang}, {Ciprini}, {Claus}, {Cognard}, {Cohen-Tanugi}, {Cominsky}, {de Palma}, {Dermer}, {do Couto e Silva}, {Dormody}, {Drell}, {Dubois}, {Dumora}, {Espinoza}, {Favuzzi}, {Fegan}, {Ferrara}, {Focke}, {Fortin}, {Fukazawa}, {Fusco}, {Gargano}, {Gasparrini}, {Gehrels}, {Germani}, {Giglietto}, {Giordano}, {Giroletti}, {Glanzman}, {Godfrey}, {Grenier}, {Grondin}, {Grove}, {Guillemot}, {Guiriec}, {Hadasch}, {Harding}, {J{\'o}hannesson}, {Johnson}, {Johnson}, {Johnston}, {Katagiri}, {Kataoka}, {Keith}, {Kerr}, {Kn{\"o}dlseder}, {Kramer}, {Kuss}, {Lande}, {Latronico}, {Lee}, {Lemoine-Goumard}, {Longo}, {Loparco}, {Lovellette}, {Lubrano}, {Lyne}, {Manchester}, {Marelli}, {Mazziotta}, {McEnery},
  {Michelson}, {Mizuno}, {Moiseev}, {Monte}, {Monzani}, {Morselli}, {Moskalenko}, {Murgia}, {Nakamori}, {Nolan}, {Norris}, {Nuss}, {Ohsugi}, {Okumura}, {Omodei}, {Orlando}, {Ozaki}, {Paneque}, {Parent}, {Pesce-Rollins}, {Pierbattista}, {Piron}, {Porter}, {Rain{\`o}}, {Ransom}, {Ray}, {Reimer}, {Reimer}, {Reposeur}, {Ritz}, {Romani}, {Roth}, {Sadrozinski}, {Saz Parkinson}, {Shannon}, {Siskind}, {Smith}, {Spinelli}, {Stappers}, {Suson}, {Takahashi}, {Tanaka}, {Tauris}, {Thayer}, {Theureau}, {Thompson}, {Thorsett}, {Tibaldo}, {Torres}, {Tosti}, {Troja}, {Vandenbroucke}, {Van Etten}, {Vasileiou}, {Venter}, {Vianello}, {Vilchez}, {Vitale}, {Waite}, {Wang}, {Wood}, {Yang}, {Ziegler}, \& {Zimmer}}]{Freire_2011}
{Freire}, P.~C.~C., {Abdo}, A.~A., {Ajello}, M., {et~al.} 2011, Science, 334, 1107, \dodoi{10.1126/science.1207141}

\bibitem[{{Freire} {et~al.}(2017){Freire}, {Ridolfi}, {Kramer}, {Jordan}, {Manchester}, {Torne}, {Sarkissian}, {Heinke}, {D'Amico}, {Camilo}, {Lorimer}, \& {Lyne}}]{Freire_2017}
{Freire}, P.~C.~C., {Ridolfi}, A., {Kramer}, M., {et~al.} 2017, \mnras, 471, 857, \dodoi{10.1093/mnras/stx1533}

\bibitem[{{Gautam} {et~al.}(2022){Gautam}, {Ridolfi}, {Freire}, {Wharton}, {Gupta}, {Ransom}, {Oswald}, {Kramer}, \& {DeCesar}}]{Gautam2022}
{Gautam}, T., {Ridolfi}, A., {Freire}, P.~C.~C., {et~al.} 2022, \aap, 664, A54, \dodoi{10.1051/0004-6361/202243062}

\bibitem[{{Gupta} {et~al.}(2017){Gupta}, {Ajithkumar}, {Kale}, {Nayak}, {Sabhapathy}, {Sureshkumar}, {Swami}, {Chengalur}, {Ghosh}, {Ishwara-Chandra}, {Joshi}, {Kanekar}, {Lal}, \& {Roy}}]{Gupta_2017}
{Gupta}, Y., {Ajithkumar}, B., {Kale}, H.~S., {et~al.} 2017, Current Science, 113, 707, \dodoi{10.18520/cs/v113/i04/707-714}

\bibitem[{{Hankins}(1971)}]{Hankins_1971}
{Hankins}, T.~H. 1971, \apj, 169, 487, \dodoi{10.1086/151164}

\bibitem[{{Harris}(1996)}]{Harris_1996}
{Harris}, W.~E. 1996, \aj, 112, 1487, \dodoi{10.1086/118116}

\bibitem[{{He} \& {Shi}(2024)}]{He_2024}
{He}, Q., \& {Shi}, X. 2024, \mnras, 527, 5183, \dodoi{10.1093/mnras/stad3561}

\bibitem[{{Hessels} {et~al.}(2007){Hessels}, {Ransom}, {Stairs}, {Kaspi}, \& {Freire}}]{Hessels_2007}
{Hessels}, J.~W.~T., {Ransom}, S.~M., {Stairs}, I.~H., {Kaspi}, V.~M., \& {Freire}, P.~C.~C. 2007, \apj, 670, 363, \dodoi{10.1086/521780}

\bibitem[{{Jia} \& {Li}(2016)}]{Jia_2016}
{Jia}, K., \& {Li}, X.-D. 2016, \apj, 830, 153, \dodoi{10.3847/0004-637X/830/2/153}

\bibitem[{{Johnson} {et~al.}(2013){Johnson}, {Guillemot}, {Kerr}, {Cognard}, {Ray}, {Wolff}, {B{\'e}gin}, {Janssen}, {Romani}, {Venter}, {Grove}, {Freire}, {Wood}, {Cheung}, {Casandjian}, {Stairs}, {Camilo}, {Espinoza}, {Ferrara}, {Harding}, {Johnston}, {Kramer}, {Lyne}, {Michelson}, {Ransom}, {Shannon}, {Smith}, {Stappers}, {Theureau}, \& {Thorsett}}]{Johnson_2013}
{Johnson}, T.~J., {Guillemot}, L., {Kerr}, M., {et~al.} 2013, \apj, 778, 106, \dodoi{10.1088/0004-637X/778/2/106}

\bibitem[{{Kale} \& {Ishwara-Chandra}(2021)}]{Kale_2021}
{Kale}, R., \& {Ishwara-Chandra}, C.~H. 2021, Experimental Astronomy, 51, 95, \dodoi{10.1007/s10686-020-09677-6}

\bibitem[{{Kudale} {et~al.}(2024){Kudale}, {Roy}, {Chengalur}, {Sharma}, \& {Kumari}}]{Kudale_2024}
{Kudale}, S., {Roy}, J., {Chengalur}, J.~N., {Sharma}, S., \& {Kumari}, S. 2024, \apj, 972, 61, \dodoi{10.3847/1538-4357/ad6315}

\bibitem[{{Kumari} {et~al.}(2024){Kumari}, {Bhattacharyya}, {Sharan}, {Johnston}, {Weltevrede}, {Stappers}, {Kansabanik}, {Roy}, \& {Ghosh}}]{Kumari_2024}
{Kumari}, S., {Bhattacharyya}, B., {Sharan}, R., {et~al.} 2024, \apj, 973, 19, \dodoi{10.3847/1538-4357/ad6145}

\bibitem[{{Li} {et~al.}(2024){Li}, {Zhang}, {Yao}, {Yin}, {Eatough}, {Li}, {Li}, {Lian}, {Pan}, {Dai}, {Li}, {Zhang}, {Su}, {Wu}, {Liu}, {Liu}, {Wang}, {Qian}, \& {Pan}}]{Li_2024}
{Li}, B., {Zhang}, L.-y., {Yao}, J., {et~al.} 2024, \apj, 972, 43, \dodoi{10.3847/1538-4357/ad5a82}

\bibitem[{{Lian} {et~al.}(2023){Lian}, {Pan}, {Zhang}, {Freire}, {Cao}, \& {Qian}}]{Lian_2023}
{Lian}, Y., {Pan}, Z., {Zhang}, H., {et~al.} 2023, \apjl, 951, L37, \dodoi{10.3847/2041-8213/acdee7}

\bibitem[{{Lian} {et~al.}(2025){Lian}, {Freire}, {Cao}, {Cadelano}, {Pallanca}, {Pan}, {Zhang}, {Li}, \& {Qian}}]{Lian_2025}
{Lian}, Y., {Freire}, P.~C.~C., {Cao}, S., {et~al.} 2025, arXiv e-prints, arXiv:2502.02042, \dodoi{10.48550/arXiv.2502.02042}

\bibitem[{{Lorimer} \& {Kramer}(2004)}]{Lorimer2004}
{Lorimer}, D.~R., \& {Kramer}, M. 2004, {Handbook of Pulsar Astronomy}, Vol.~4

\bibitem[{{Lynch} {et~al.}(2012){Lynch}, {Freire}, {Ransom}, \& {Jacoby}}]{Lynch2012}
{Lynch}, R.~S., {Freire}, P. C.~C., {Ransom}, S.~M., \& {Jacoby}, B.~A. 2012, \apj, 745, 109, \dodoi{10.1088/0004-637X/745/2/109}

\bibitem[{{Lynch} {et~al.}(2011){Lynch}, {Ransom}, {Freire}, \& {Stairs}}]{Lynch2011}
{Lynch}, R.~S., {Ransom}, S.~M., {Freire}, P. C.~C., \& {Stairs}, I.~H. 2011, \apj, 734, 89, \dodoi{10.1088/0004-637X/734/2/89}

\bibitem[{{Manchester} {et~al.}(2005){Manchester}, {Hobbs}, {Teoh}, \& {Hobbs}}]{Manchester_2025}
{Manchester}, R.~N., {Hobbs}, G.~B., {Teoh}, A., \& {Hobbs}, M. 2005, \aj, 129, 1993, \dodoi{10.1086/428488}

\bibitem[{{McKinnon}(2014)}]{McKinnon_2014}
{McKinnon}, M.~M. 2014, \pasp, 126, 476, \dodoi{10.1086/676975}

\bibitem[{{McMillan}(2017)}]{McMillan2017}
{McMillan}, P.~J. 2017, \mnras, 465, 76, \dodoi{10.1093/mnras/stw2759}

\bibitem[{{Padmanabh} {et~al.}(2024){Padmanabh}, {Ransom}, {Freire}, {Ridolfi}, {Taylor}, {Choza}, {Clark}, {Abbate}, {Bailes}, {Barr}, {Buchner}, {Burgay}, {DeCesar}, {Chen}, {Corongiu}, {Champion}, {Dutta}, {Geyer}, {Hessels}, {Kramer}, {Possenti}, {Stairs}, {Stappers}, {Venkatraman Krishnan}, {Vleeschower}, \& {Zhang}}]{Padmanabh_2024}
{Padmanabh}, P.~V., {Ransom}, S.~M., {Freire}, P.~C.~C., {et~al.} 2024, \aap, 686, A166, \dodoi{10.1051/0004-6361/202449303}

\bibitem[{{Pan} {et~al.}(2021{\natexlab{a}}){Pan}, {Qian}, {Ma}, {Liu}, {Wang}, {Luo}, {Yan}, {Ransom}, {Lorimer}, {Li}, \& {Jiang}}]{Pan2021a}
{Pan}, Z., {Qian}, L., {Ma}, X., {et~al.} 2021{\natexlab{a}}, \apjl, 915, L28, \dodoi{10.3847/2041-8213/ac0bbd}

\bibitem[{{Pan} {et~al.}(2021{\natexlab{b}}){Pan}, {Ma}, {Qian}, {Wang}, {Yan}, {Luo}, {Ransom}, {Lorimer}, \& {Jiang}}]{Pan2021b}
{Pan}, Z., {Ma}, X.-Y., {Qian}, L., {et~al.} 2021{\natexlab{b}}, Research in Astronomy and Astrophysics, 21, 143, \dodoi{10.1088/1674-4527/21/6/143}

\bibitem[{{Park} {et~al.}(2021){Park}, {Folkner}, {Williams}, \& {Boggs}}]{Park_2021}
{Park}, R.~S., {Folkner}, W.~M., {Williams}, J.~G., \& {Boggs}, D.~H. 2021, \aj, 161, 105, \dodoi{10.3847/1538-3881/abd414}

\bibitem[{{Phinney}(1992)}]{Phinney_1992}
{Phinney}, E.~S. 1992, Philosophical Transactions of the Royal Society of London Series A, 341, 39, \dodoi{10.1098/rsta.1992.0084}

\bibitem[{{Pooley} {et~al.}(2003){Pooley}, {Lewin}, {Anderson}, {Baumgardt}, {Filippenko}, {Gaensler}, {Homer}, {Hut}, {Kaspi}, {Makino}, {Margon}, {McMillan}, {Portegies Zwart}, {van der Klis}, \& {Verbunt}}]{Pooley_2003}
{Pooley}, D., {Lewin}, W. H.~G., {Anderson}, S.~F., {et~al.} 2003, \apjl, 591, L131, \dodoi{10.1086/377074}

\bibitem[{{Possenti} {et~al.}(2005){Possenti}, {D'Amico}, {Corongiu}, {Manchester}, {Sarkissian}, {Camilo}, \& {Lyne}}]{Possenti2005}
{Possenti}, A., {D'Amico}, N., {Corongiu}, A., {et~al.} 2005, in Astronomical Society of the Pacific Conference Series, Vol. 328, Binary Radio Pulsars, ed. F.~A. {Rasio} \& I.~H. {Stairs}, 189

\bibitem[{{Possenti} {et~al.}(2001){Possenti}, {D'Amico}, {Manchester}, {Sarkissian}, {Lyne}, \& {Camilo}}]{Possenti2001}
{Possenti}, A., {D'Amico}, N., {Manchester}, R.~N., {et~al.} 2001, arXiv e-prints, astro, \dodoi{10.48550/arXiv.astro-ph/0108343}

\bibitem[{{Pr{\v{s}}a} {et~al.}(2016){Pr{\v{s}}a}, {Harmanec}, {Torres}, {Mamajek}, {Asplund}, {Capitaine}, {Christensen-Dalsgaard}, {Depagne}, {Haberreiter}, {Hekker}, {Hilton}, {Kopp}, {Kostov}, {Kurtz}, {Laskar}, {Mason}, {Milone}, {Montgomery}, {Richards}, {Schmutz}, {Schou}, \& {Stewart}}]{Prsa_2016}
{Pr{\v{s}}a}, A., {Harmanec}, P., {Torres}, G., {et~al.} 2016, \aj, 152, 41, \dodoi{10.3847/0004-6256/152/2/41}

\bibitem[{{Ransom} {et~al.}(2005){Ransom}, {Hessels}, {Stairs}, {Freire}, {Camilo}, {Kaspi}, \& {Kaplan}}]{Ransom2005}
{Ransom}, S.~M., {Hessels}, J. W.~T., {Stairs}, I.~H., {et~al.} 2005, Science, 307, 892, \dodoi{10.1126/science.1108632}

\bibitem[{{Ransom} {et~al.}(2004){Ransom}, {Stairs}, {Backer}, {Greenhill}, {Bassa}, {Hessels}, \& {Kaspi}}]{Ransom2004}
{Ransom}, S.~M., {Stairs}, I.~H., {Backer}, D.~C., {et~al.} 2004, \apj, 604, 328, \dodoi{10.1086/381730}

\bibitem[{{Reddy} {et~al.}(2017){Reddy}, {Kudale}, {Gokhale}, {Halagalli}, {Raskar}, {de}, {Gnanaraj}, {Ajith Kumar}, \& {Gupta}}]{Reddy_2017}
{Reddy}, S.~H., {Kudale}, S., {Gokhale}, U., {et~al.} 2017, Journal of Astronomical Instrumentation, 6, 1641011, \dodoi{10.1142/S2251171716410117}

\bibitem[{{Ridolfi} {et~al.}(2021){Ridolfi}, {Gautam}, {Freire}, {Ransom}, {Buchner}, {Possenti}, {Venkatraman Krishnan}, {Bailes}, {Kramer}, {Stappers}, {Abbate}, {Barr}, {Burgay}, {Camilo}, {Corongiu}, {Jameson}, {Padmanabh}, {Vleeschower}, {Champion}, {Chen}, {Geyer}, {Karastergiou}, {Karuppusamy}, {Parthasarathy}, {Reardon}, {Serylak}, {Shannon}, \& {Spiewak}}]{Ridolfi2021}
{Ridolfi}, A., {Gautam}, T., {Freire}, P.~C.~C., {et~al.} 2021, \mnras, 504, 1407, \dodoi{10.1093/mnras/stab790}

\bibitem[{{Ridolfi} {et~al.}(2022{\natexlab{a}}){Ridolfi}, {Freire}, {Gautam}, {Ransom}, {Barr}, {Buchner}, {Burgay}, {Abbate}, {Venkatraman Krishnan}, {Vleeschower}, {Possenti}, {Stappers}, {Kramer}, {Chen}, {Padmanabh}, {Champion}, {Bailes}, {Levin}, {Keane}, {Breton}, {Bezuidenhout}, {Grie{\ss}meier}, {K{\"u}nkel}, {Men}, {Camilo}, {Geyer}, {Hugo}, {Jameson}, {Parthasarathy}, \& {Serylak}}]{Ridolfi2022}
{Ridolfi}, A., {Freire}, P.~C.~C., {Gautam}, T., {et~al.} 2022{\natexlab{a}}, \aap, 664, A27, \dodoi{10.1051/0004-6361/202143006}

\bibitem[{{Ridolfi} {et~al.}(2022{\natexlab{b}}){Ridolfi}, {Freire}, {Gautam}, {Ransom}, {Barr}, {Buchner}, {Burgay}, {Abbate}, {Venkatraman Krishnan}, {Vleeschower}, {Possenti}, {Stappers}, {Kramer}, {Chen}, {Padmanabh}, {Champion}, {Bailes}, {Levin}, {Keane}, {Breton}, {Bezuidenhout}, {Grie{\ss}meier}, {K{\"u}nkel}, {Men}, {Camilo}, {Geyer}, {Hugo}, {Jameson}, {Parthasarathy}, \& {Serylak}}]{Ridolfi_2022}
---. 2022{\natexlab{b}}, \aap, 664, A27, \dodoi{10.1051/0004-6361/202143006}

\bibitem[{{Sarazin} {et~al.}(2003){Sarazin}, {Kundu}, {Irwin}, {Sivakoff}, {Blanton}, \& {Randall}}]{Sarazin_2003}
{Sarazin}, C.~L., {Kundu}, A., {Irwin}, J.~A., {et~al.} 2003, \apj, 595, 743, \dodoi{10.1086/377467}

\bibitem[{{Shklovskii}(1970)}]{Shklovskii_1970}
{Shklovskii}, I.~S. 1970, \sovast, 13, 562

\bibitem[{{Singleton} {et~al.}(2024){Singleton}, {DeCesar}, {Dai}, {Bhakta}, {Ransom}, {Strader}, {Chomiuk}, \& {Miller-Jones}}]{Singleton_2024}
{Singleton}, J., {DeCesar}, M., {Dai}, S., {et~al.} 2024, arXiv e-prints, arXiv:2412.11271, \dodoi{10.48550/arXiv.2412.11271}

\bibitem[{{Splaver} {et~al.}(2002){Splaver}, {Nice}, {Arzoumanian}, {Camilo}, {Lyne}, \& {Stairs}}]{Splaver_2002}
{Splaver}, E.~M., {Nice}, D.~J., {Arzoumanian}, Z., {et~al.} 2002, \apj, 581, 509, \dodoi{10.1086/344202}

\bibitem[{Swarup {et~al.}(1991)Swarup, Ananthakrishnan, Kapahi, Rao, Subrahmanya, \& Kulkarni}]{Swarup_1991}
Swarup, G., Ananthakrishnan, S., Kapahi, V.~K., {et~al.} 1991, Current Science, 60, 95.
\newblock \url{http://www.jstor.org/stable/24094934}

\bibitem[{{Tam} {et~al.}(2016){Tam}, {Hui}, \& {Kong}}]{Tam_2016}
{Tam}, P.-H.~T., {Hui}, C.~Y., \& {Kong}, A. K.~H. 2016, Journal of Astronomy and Space Sciences, 33, 1, \dodoi{10.5140/JASS.2016.33.1.1}

\bibitem[{{Tam} {et~al.}(2011){Tam}, {Kong}, {Hui}, {Cheng}, {Li}, \& {Lu}}]{Tam_2011}
{Tam}, P.~H.~T., {Kong}, A.~K.~H., {Hui}, C.~Y., {et~al.} 2011, \apj, 729, 90, \dodoi{10.1088/0004-637X/729/2/90}

\bibitem[{{Tauris} \& {Savonije}(1999)}]{Tauris_1999}
{Tauris}, T.~M., \& {Savonije}, G.~J. 1999, \aap, 350, 928, \dodoi{10.48550/arXiv.astro-ph/9909147}

\bibitem[{{Tauris} \& {van den Heuvel}(2023)}]{Tauris_2023}
{Tauris}, T.~M., \& {van den Heuvel}, E. P.~J. 2023, {Physics of Binary Star Evolution. From Stars to X-ray Binaries and Gravitational Wave Sources}, \dodoi{10.48550/arXiv.2305.09388}

\bibitem[{{Taylor} \& {Weisberg}(1989)}]{Taylor_1989}
{Taylor}, J.~H., \& {Weisberg}, J.~M. 1989, \apj, 345, 434, \dodoi{10.1086/167917}

\bibitem[{{Vasiliev} \& {Baumgardt}(2021)}]{Vasiliev_2021}
{Vasiliev}, E., \& {Baumgardt}, H. 2021, \mnras, 505, 5978, \dodoi{10.1093/mnras/stab1475}

\bibitem[{{Verbunt} \& {Freire}(2014)}]{Verbunt_Freire_2014}
{Verbunt}, F., \& {Freire}, P. C.~C. 2014, \aap, 561, A11, \dodoi{10.1051/0004-6361/201321177}

\bibitem[{{Verbunt} \& {Hut}(1987)}]{Verbunt_Hut_1987}
{Verbunt}, F., \& {Hut}, P. 1987, in IAU Symposium, Vol. 125, The Origin and Evolution of Neutron Stars, ed. D.~J. {Helfand} \& J.~H. {Huang}, 187

\bibitem[{{Vleeschower} {et~al.}(2024){Vleeschower}, {Corongiu}, {Stappers}, {Freire}, {Ridolfi}, {Abbate}, {Ransom}, {Possenti}, {Padmanabh}, {Balakrishnan}, {Kramer}, {Venkatraman Krishnan}, {Zhang}, {Bailes}, {Barr}, {Buchner}, \& {Chen}}]{Vleeschower_2024}
{Vleeschower}, L., {Corongiu}, A., {Stappers}, B.~W., {et~al.} 2024, \mnras, 530, 1436, \dodoi{10.1093/mnras/stae816}

\bibitem[{{Wang} {et~al.}(2020){Wang}, {Peng}, {Stappers}, {Liu}, {Keith}, {Lyne}, {Lu}, {Yu}, {Kou}, {Yan}, {Jiang}, {Jin}, {Li}, {Li}, {Qian}, {Wang}, {Yue}, {Zhang}, {Zhang}, {Zhu}, \& {FAST Collaboration}}]{Wang_2020}
{Wang}, L., {Peng}, B., {Stappers}, B.~W., {et~al.} 2020, \apj, 892, 43, \dodoi{10.3847/1538-4357/ab76cc}

\bibitem[{{Webb} {et~al.}(2006){Webb}, {Wheatley}, \& {Barret}}]{Webb_2006}
{Webb}, N.~A., {Wheatley}, P.~J., \& {Barret}, D. 2006, \aap, 445, 155, \dodoi{10.1051/0004-6361:20053010}

\bibitem[{{Wu} {et~al.}(2024){Wu}, {Pan}, {Qian}, {Ransom}, {Eatough}, {Wang}, {Freire}, {Liu}, {Yan}, {Luo}, {Zhang}, {Li}, {Yin}, {Li}, {Li}, {Dai}, {Li}, {Zhang}, {Liu}, \& {Pan}}]{Wu_2024}
{Wu}, Y., {Pan}, Z., {Qian}, L., {et~al.} 2024, \apjl, 974, L23, \dodoi{10.3847/2041-8213/ad7b9e}

\bibitem[{{Yan} {et~al.}(2021){Yan}, {Pan}, {Ransom}, {Lorimer}, {Qian}, {Wang}, {Shen}, {Li}, {Jiang}, {Luo}, {Liu}, \& {Huang}}]{Yan_2021}
{Yan}, Z., {Pan}, Z.-c., {Ransom}, S.~M., {et~al.} 2021, \apj, 921, 120, \dodoi{10.3847/1538-4357/ac25eb}

\bibitem[{{Yao} {et~al.}(2017){Yao}, {Manchester}, \& {Wang}}]{Yao_2017}
{Yao}, J.~M., {Manchester}, R.~N., \& {Wang}, N. 2017, \apj, 835, 29, \dodoi{10.3847/1538-4357/835/1/29}

\bibitem[{{Yin} {et~al.}(2024){Yin}, {Zhang}, {Qian}, {Eatough}, {Li}, {Lorimer}, {Dai}, {Li}, {Zhang}, {Li}, {Su}, {Wu}, {Pan}, {Lian}, {Liu}, {Yan}, \& {Pan}}]{Yin_2024}
{Yin}, D., {Zhang}, L.-y., {Qian}, L., {et~al.} 2024, \apjl, 969, L7, \dodoi{10.3847/2041-8213/ad534e}

\bibitem[{{Zhang} \& {Cheng}(2003)}]{Zhang_2003}
{Zhang}, L., \& {Cheng}, K.~S. 2003, \aap, 398, 639, \dodoi{10.1051/0004-6361:20021570}

\bibitem[{{Zhang} {et~al.}(2023){Zhang}, {Freire}, {Ridolfi}, {Pan}, {Zhao}, {Heinke}, {Chen}, {Cadelano}, {Pallanca}, {Hou}, {Fu}, {Dai}, {G{\"u}gercino{\u{g}}lu}, {Guo}, {Hessels}, {Hu}, {Li}, {Ni}, {Pan}, {Ransom}, {Ruan}, {Stairs}, {Tsai}, {Wang}, {Wang}, {Wang}, {Wu}, {Yuan}, {Zhang}, {Zhu}, {Zhang}, \& {Li}}]{Zhang_2023}
{Zhang}, L., {Freire}, P. C.~C., {Ridolfi}, A., {et~al.} 2023, \apjs, 269, 56, \dodoi{10.3847/1538-4365/acfb03}

\bibitem[{{Zhang} {et~al.}(2022){Zhang}, {Xing}, \& {Wang}}]{Zhang_2022}
{Zhang}, P., {Xing}, Y., \& {Wang}, Z. 2022, \apjl, 935, L36, \dodoi{10.3847/2041-8213/ac88bf}

\bibitem[{{Zhou} {et~al.}(2024){Zhou}, {Wang}, {Li}, {Fang}, {Miao}, {Freire}, {Zhang}, {Zhang}, {Chen}, {Feng}, {Xiao}, {Xie}, {Zhang}, {Jin}, {Wang}, {Ke}, {Guo}, {Zhao}, {Niu}, {Zhu}, {Xue}, {Wang}, {Wu}, {Gan}, {Sun}, {Wang}, {Zhang}, {Zhang}, {Cao}, \& {Lu}}]{Zhou_2024}
{Zhou}, D., {Wang}, P., {Li}, D., {et~al.} 2024, Science China Physics, Mechanics, and Astronomy, 67, 269512, \dodoi{10.1007/s11433-023-2362-x}

\end{thebibliography}
\bibliographystyle{aasjournal}

\end{document}